\documentclass[useAMS,usenatbib]{mn2e} 
\usepackage{epsfig} \usepackage{amsmath} \usepackage{rotating}

\newcommand{\be}{\begin{equation}} \newcommand{\beq}{\begin{equation}}
\newcommand{\ba}{\begin{eqnarray}} \newcommand{\ee}{\end{equation}}
\newcommand{\eeq}{\end{equation}} \newcommand{\ea}{\end{eqnarray}}
  
\newcommand{\hs}{\hspace{1mm}} 
  
  \newcommand{\apj}{ApJ}
\newcommand{\aap}{A\&A} \newcommand{\apjl}{ApJL}
\newcommand{\mnras}{MNRAS} \newcommand{\aj}{AJ}
\newcommand{\apjs}{ApJS} \newcommand{\nat}{{\it Nature}}
 
\newcommand{\apss}{APSS} 
\newcommand{\physrep}{PhR}

\def\lsim{~\rlap{$<$}{\lower 1.0ex\hbox{$\sim$}}}

\def\gsim{~\rlap{$>$}{\lower 1.0ex\hbox{$\sim$}}}

\title[Ly$\alpha$ Blobs from Cold Accretion]{Ly$\alpha$ Blobs as an
Observational Signature of Cold Accretion Streams into Galaxies}

\author[Dem Style]{Dem$^{1}$\thanks{E-mail:mdijkstr@cfa.harvard.edu}, }

\author[Dijkstra \& Loeb]{Mark
Dijkstra\thanks{E-mail:mdijkstr@cfa.harvard.edu} and Abraham Loeb\\
$^{1}$Astronomy Department, Harvard University, 60 Garden Street,
Cambridge, MA 02138, USA}

\begin{document}

\date{\today} \pagerange{\pageref{firstpage}--\pageref{lastpage}}
\pubyear{2006}

\maketitle

\label{firstpage}
\begin{abstract}
Recent hydrodynamic simulations of galaxy formation reveal streams of
cold ($T\sim 10^4$ K) gas flowing into the centers of dark matter
halos as massive as $10^{12-13.5}M_{\odot}$ at redshifts $z\sim
1$--$3$. In this paper we show that if $\gsim 20\%$ of the gravitational binding energy of the
gas is radiated away, then the simulated cold flows are spatially
extended Ly$\alpha$ sources with luminosities, Ly$\alpha$ line widths,
and number densities that are comparable to those of observed
Ly$\alpha$ blobs. Furthermore, the filamentary structure of the cold
flows can explain the wide range of observed Ly$\alpha$ blob
morphologies. Since the most massive halos form in dense environments,
the association of Ly$\alpha$ blobs with overdense regions arise
naturally. We argue that Ly$\alpha$ blobs - even those which are
clearly associated with starburst galaxies or quasars - provide direct
observational support for the cold accretion mode of galaxies. We
discuss various testable predictions of this association.
\end{abstract}

 
\section{Introduction}
\label{sec:intro}
The physical origin of spatially extended Ly$\alpha$ sources, also
known as Ly$\alpha$ blobs (LABs), is still enatic. LABs have been
associated with cooling radiation, in which gas that collapses inside
a host dark matter halo releases a significant fraction of its
gravitational binding energy in Ly$\alpha$ line emission \citep[also
see Katz \& Gunn 1991]{HS00,Fardal01,Birnboim03,D06}. Other mechanism
that have been invoked to explain the origin of LABs include:
photoionization of cold ($T\sim 10^4$ K), dense, spatially extended
gas by obscured quasars \citep{HR01,Geach09},  population III stars
\citep{JH06}, or spatially extended inverse Compton X-ray emission
\citep{Scharf03}; the compression of ambient gas by superwinds into
dense, cold Ly$\alpha$ emitting shells
\citep[e.g.][]{Taniguchi00,Mori04}; or star formation that is
triggered by relativistic jets \citep{Rees89}; or some combination of
photoionization and cooling \citep{Fur05}.

Since their discovery by Steidel et al. (2000, also see Keel et al. 1999), several tens of new LABs have been found
\citep[e.g.][]{Matsuda04,Dey05,Saito06,Smith08b}. These are typically
associated with massive halos that reside in dense parts of the
Universe \citep{Steidel00,Matsuda04,Matsuda06}. Multi-wavelength
studies of Ly$\alpha$ blobs have revealed a clear association of the
brighter blobs with regular Lyman Break galaxies (LBGs, e.g. Matsuda
et al. 2004), sub-millimeter (sub-mm) and infrared (IR) sources which
imply star formation rates of $\sim 10^3M_{\odot}$ yr$^{-1}$
\citep{Chapman01,Geach05,Geach07,Matsuda07}, or with unobscured
\citep{Bunker03,Weidinger04} and obscured quasars
\citep{Basu04,Geach07,Smith08b}.  However, in other blobs this
association has been ruled out, which has led to the conclusion that
cooling radiation by cold accreting gas may have been observed
\citep{Nilsson06,Matsuda06,Saito08,Smith07,Smith08a}.

We propose that the sources that are associated with the Ly$\alpha$
blobs may have little to do in powering the spatially extended
Ly$\alpha$ emission. This proposal is physically motivated by recent
hydrodynamical simulations of galaxy formation, which show that
baryons assemble into galaxies through a two-phase medium which
contains filamentary streams of cold ($T\sim 10^4$ K) gas embedded
within a hot gaseous halo
\citep[e.g][]{Keres05,BD06,Ocvirk08,Dekel08,Keres08,Agertz09}. The
cold flows contain $\sim 5$--$50\%$ of the total gas content
\citep{B07,Keres08} in halos with masses in the range $M_{\rm halo}
\sim 10^{12}$--$10^{13.5}M_{\odot}$ \citep{Dekel08}. As we will show
in this paper, these cold streams are probable sources of spatially
extended Ly$\alpha$ emission. Under very reasonable assumptions we
find that the cold accretion model 'predicts' the existence of
spatially extended Ly$\alpha$ sources with properties reminiscent of
the observed LABs around massive halos ($M_{\rm halo}\gsim
10^{12}M_{\odot}$).

The outline of this paper is as follows: we describe our model in
\S~\ref{sec:model}. In \S~\ref{sec:results} we present our results,
before discussing our work in \S~\ref{sec:discuss} and presenting our
conclusions in \S~\ref{sec:conc}. The cosmological parameter values
used throughout our discussion are
$(\Omega_m,\Omega_{\Lambda},\Omega_b,h,\sigma_8)=(0.27,0.73,0.046,0.70,0.82)$
\citep{Komatsu08}, and we denote the primordial helium abundance by
mass as $Y_{\rm He}=0.24$ \citep[e.g][]{Helium}.

\section{The Model}
\label{sec:model}

Our model is based on the scenario for cold accretion that has emerged
from recent simulations \citep{Keres08}. 'Hot' gas that is in
hydrostatic equilibrium with the dark matter potential well at the
virial temperature of the dark matter halo is in pressure equilibrium
with the cold gas \citep[also see][]{FR85,Rees89}, which makes up a
fraction $f_{\rm cold}$ of the total gas mass of the halo, $M_{\rm
gas}$.  We further assume that the gas mass fraction, $f_{\rm
gas}$, is half of the universal value. Hence, the total gas mass
$M_{\rm gas}\equiv f_{\rm gas}M_{\rm halo}=0.5\Omega_b/\Omega_{\rm
M}M_{\rm halo}=0.08M_{\rm halo}$. This accounts for a substantial
fraction of baryons that may be locked up in (dwarf) galaxies that
reside within the halo of interest. The assumed gas mass fraction
$f_{\rm gas}=0.08$ is in good agreement with the value derived for
groups of galaxies at $z < 1$ \citep[e.g.][]{fbar}.  As the cold
streams navigate to the center of their host dark matter halo, they
are progressively heated by the release of a fraction of their
gravitational potential energy through weak shocks
\citep{RO77,HR01}. This heating of the cold gas is balanced by
radiative cooling , mostly through the Ly$\alpha$ emission line. Next,
we describe the model in more detail.

\begin{itemize}
\item We use the gas density profile for hot gas that is in
hydrostatic equilibrium with an NFW \citep{NFW} dark matter potential
well with a concentration parameter $c=3.8$ \citep{Gao08} at the
virial temperature of the halo as derived by Makino et al (1998, their
Eqs.~8 and~11). This profile is very similar to an isothermal
$\beta$-model, which -- according to the simulations -- provides an
accurate description of the hot gas component \citep{Keres08}.  
In \S~\ref{sec:unc} we show that our final results are robust against
variations in the assumed density profile. 

\item The hot gas reaches the virial temperature, $T_{\rm h}(r)=T_{\rm
vir}=1.9\times 10^6 {\rm K} (M_{\rm halo}/10^{12} M_\odot)^{2/3}
[(1+z)/4]$, and co-exists in pressure equilibrium with the cold
gas. Pressure equilibrium between the hot (`h') and cold (`c') gas
implies that $n_{\rm c}(r)T_{\rm c}(r)=n_{\rm h}(r)T_{\rm h}(r)$, in
which $n(r)$ and $T(r)$ denote the number density of particles and the
gas temperature, respectively. We obtain $T_c(r)$ under the assumption
that cooling balances the heating of the cold flows that occurs as
they navigate into the center of the dark matter halo.

We parametrize the gravitational heating rate by assuming that a
fraction $f_{\rm grav}$ ($f_{\rm hot}$) of the change in the
gravitational potential energy of each gas element along its
trajectory is converted into heat in the cold (hot) gas. The
remaining fraction $(1-f_{\rm grav}-f_{\rm hot})$ is converted into
additional kinetic energy of the gas element. Throughout the
paper, we assume that the transfer of energy into the hot gas is
negligible, i.e. $f_{\rm hot}=0$. This assumption appears reasonable
given that the majority of the cold gas mass is in smooth
continuous streams \citep[][]{Keres08,Dekel08}, as opposed
to discrete clouds which are likely to heat the hot gas
\citep[e.g.][and references therein]{BD08}. Under this assumption
, which is discussed in more detail in \S~\ref{sec:hot},
$f_{\rm grav}=1$ corresponds to infall at a constant velocity, while
$f_{\rm grav}=0$ corresponds to free-fall.

For the sake of simplicity, we adopt the conservative working
assumption of a constant non-zero $f_{\rm grav}$ inside the virial
radius ($r_{\rm vir}$) of galaxy halos and $f_{\rm grav}=0$
outside. The most recent Smoothed Particle Hydrodynamical (SPH)
simulations \citep{Keres08} indicate that the cold flows propagate
inward at an approximately constant speed, implying $f_{\rm grav} \sim
1$; however, Adaptive Mesh Refinement (AMR) simulations
\citep{Ocvirk08,Dekel08} indicate that the cold gas accelerates
throughout its motion and that therefore $f_{\rm grav}$ may be smaller
\footnote{Our discussion focuses on the extended Ly$\alpha$ emission,
and not on the compact core of the galaxy where the cold streams are
finally brought to rest. Note that dust may suppress the Ly$\alpha$
luminosity from the core, but is less likely to affect the cold
streams which carry metal-poor intergalactic gas.}. Future simulations
might be able to refine our working assumption by resolving the
precise dynamics and heating of the cold flows.  Hence, the heating
rate per particle is given by
\begin{equation}
H(r)=f_{\rm grav} \times \frac{GM(<r)\mu m_p}{r^2}v(r)+H_{\gamma}(r),
\label{eq:heat}
\end{equation} 
where $\mu$ is the mean molecular weight per particle in the cold flow
(in units of $m_p$), and $v(r)$ denotes the infall velocity which is
given
\begin{equation}
v^2(r)=2v^2(r_{\rm vir})+2(1-f_{\rm grav}-f_{\rm hot})\int_{r_{\rm
vir}}^r ds\frac{GM(<s)}{s^2},
\label{eq:vel}
\end{equation} 
 and we assume that $v(r_{\rm vir})=\sqrt{2}v_{\rm circ}$ (with our
results not being sensitive to this choice). The term $H_{\gamma}(r)$
denotes the heating rate due to absorption of ionizing radiation. At
the typical densities in the cold flows ($n_c\gsim 0.1$ cm$^{-3}$) the
gas is self-shielded from external ionizing radiation, and hence
$H_{\gamma}(r)=0$.  It is possible however, that galaxies embedded
within the cold flow may photoionize some surrounding region which
would locally boost $H(r)$ \citep{Fur05}. In any case, ignoring this
extra photoheating term only makes our predicted Ly$\alpha$
luminosities from the cold flows smaller, and therefore makes our
results more conservative. As was mentioned above, we
assume that $f_{\rm hot}=0$.

We obtain an equilibrium temperature $T_c(r)$ at radius $r$ by
equating $H(r)$ to the cooling rate per particle in the cold flow,
which is given by,
\begin{equation}
\Lambda(r,T,x_{\rm HI})={1\over n_{\rm c}} \left[n_e n_{\rm HI} C_{\rm
cool}(T_{\rm c})+n_e n_{\rm HII} R_{\rm cool}(T_{\rm c})\right] .
\end{equation}
Here, the first term in the square brackets denotes the cooling rate
(in erg s$^{-1}$ cm$^{-3}$) due to collisional excitation of H atoms
by electrons. The second term denotes the cooling rate due to
recombination events of free electrons and protons (other cooling
processes are negligibly small for the typical gas temperature in the
cold flow). The rate coefficients $C_{\rm cool}(T_{\rm c})$ and
$R_{\rm cool}(T_{\rm c})$ were taken from Hui \& Gnedin (1997,
hereafter HG97). Finally, the ionization state of the gas determines
its cooling rate. Under the assumption that the gas is self-shielding
(which is justified later) we obtain a one-to-one relation between $T$
and $x_{\rm HI}$ through $x_{\rm HI}=\alpha_{\rm rec,B}(T_{\rm
c})/(\alpha_{\rm rec,B}(T_{\rm c})+C_{\rm ion}(T_{\rm c}))$. Here,
$C_{\rm ion}(T_{\rm c})$ denotes the collisional ionization rate
coefficient, and $\alpha_{\rm rec,B}(T_{\rm c})$ denotes the case-B
recombination coefficient (as the cold neutral gas is optically thick
in all Lyman-series lines and case-B applies; the related coefficients
were taken from HG97). At the temperatures of interest, helium is
neutral inside the cold gas and free electrons are only supplied by
hydrogen, i.e. $n_{\rm HI}(r)=x_{\rm HI}(r)n_{\rm c}(r)/[1+(Y_{\rm
He}/4)]$ and $n_{\rm HII}(r)=n_{\rm e}(r)=[1-x_{\rm HI}(r)]n_{\rm
c}(r)/[1+(Y_{\rm He}/4)]$.

In practice we assume that $T_{\rm c}=10^4$ K and compute
$n_{\rm c}$ assuming pressure equilibrium. Since temperature determines
 the ionization state of the gas in self-shielded gas, we obtain a cooling rate
which we compare to the heating rate. If $\Lambda(r,T,x_{\rm
HI})>H(r)$ ($\Lambda(r,T,x_{\rm HI})<H(r)$), then $T_{\rm c}$ is
lowered (increased) until the cooling and heating rates are equal to
within $1\%$.

\item Once the temperature, density and ionization state of the gas
have been determined, we compute the Ly$\alpha$ emissivity (in erg
s$^{-1}$ cm$^{-3}$) as a function of radius,
\begin{equation} 
\epsilon_{{\rm Ly}\alpha}(r)=n_e n_{\rm HI} C_{{\rm Ly}\alpha}(T_{\rm
c})+0.68h\nu_{\alpha}n_e n_{\rm HII} \alpha_{\rm rec,B}(T_{\rm c}).
\end{equation}
Here, the first term denotes the luminosity density in Ly$\alpha$
photons (in erg s$^{-1}$ cm$^{-3}$) following collisional excitation
of H atoms by free electrons. The collisional excitation coefficient
is given by $C_{{\rm Ly}\alpha}=3.7 \times
10^{-17}\exp(-h\nu_{\alpha}/kT)/T^{1/2}$ erg s$^{-1}$ cm$^{3}$
\citep[][p 55]{Os89}. The second term denotes the luminosity density
in Ly$\alpha$ photons following case-B recombination, and that $\sim
68\%$ of all case-B recombination events result \citep{Os89} in a
Ly$\alpha$ photon of energy $h\nu_{\alpha}=10.2$ eV. In practice, the
recombination term can be safely ignored.

\item Lastly, we obtain the total Ly$\alpha$ luminosity by integrating
over volume, namely
\begin{equation}
L_{{\rm Ly}\alpha}=\mathcal{T}_{\alpha} \int_0^{\rm vir}dr \hs 4\pi
r^2 \epsilon_{{\rm Ly}\alpha}(r)f_{\rm cold}\frac{T_{\rm c}(r)}{T_{\rm
h}}.
\end{equation}
Emission from $r > r_{\rm vir}$ is expected to have a surface
brightness that is well below the detection threshold of existing
observations, because beyond $r_{\rm vir}$ no rarefied hot gas exists
to confine the cold flow \citep{Keres08,Dekel08}, and so the cold gas
density declines considerably. The gas at $r>r_{\rm vir}$ therefore
provides a negligible contribution to the total Ly$\alpha$
luminosity. The fraction $f_{\rm cold}$ (which we assume to be
independent of radius) is taken from Figure~3 of \citet{Keres08}, 
which we parametrize as $f_{\rm cold}=0.25(M_{\rm
halo}/10^{12}M_{\odot})^{-0.55}$ for $M_{\rm halo}\leq 4\times 10^{13}
M_{\odot}$, and $f_{\rm cold}=0$ otherwise.  Furthermore,
$\mathcal{T}_{\alpha}$ denotes the fraction of Ly$\alpha$ that makes
it to the observer. The factor $f_{\rm cold}T_{\rm c}(r)/T_{\rm h}$ in
the integrand denotes the fraction of the volume that is occupied by
the cold gas in the range of radii within $r \pm dr/2$.
\end{itemize}

\section{Results}
\label{sec:results}
\begin{figure}
\vbox{\centerline{\epsfig{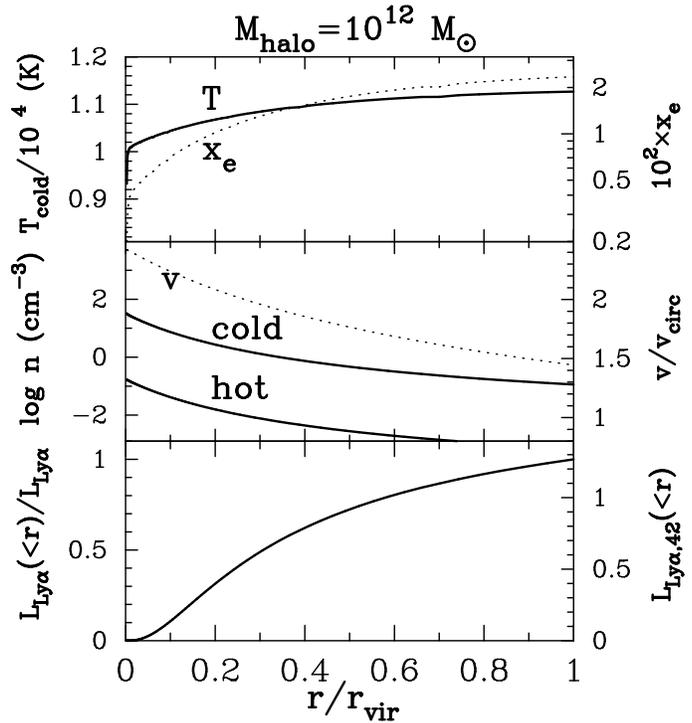}}}
\caption[]{{\it Top panel}: the {\it solid} and {\it dotted} lines
show the radial dependence of the temperature and electron fraction by
number of the cold gas phase for a dark matter halo of mass $M_{\rm
halo}=10^{12} M_{\odot}$, and under the assumption that a fraction
$f_{\rm grav}=0.3$ of the gravitational binding energy of the gas is
converted into heat. {\it Central panel}: the {\it solid} lines show
the number densities of gas particles in the cold and hot phase and
and the {\it dotted} line shows the infall velocity of the gas
expressed in units of $v_{\rm circ}$. {\it Bottom panel}: the fraction
of Ly$\alpha$ luminosity that is emitted interior to a radius $r$
(normalized by the total, $L_{\alpha,{\rm tot}}\approx 1.3 \times
10^{42}$ erg s$^{-1}$). }
\label{fig:profile}
\end{figure}
\begin{figure}
\vbox{\centerline{\epsfig{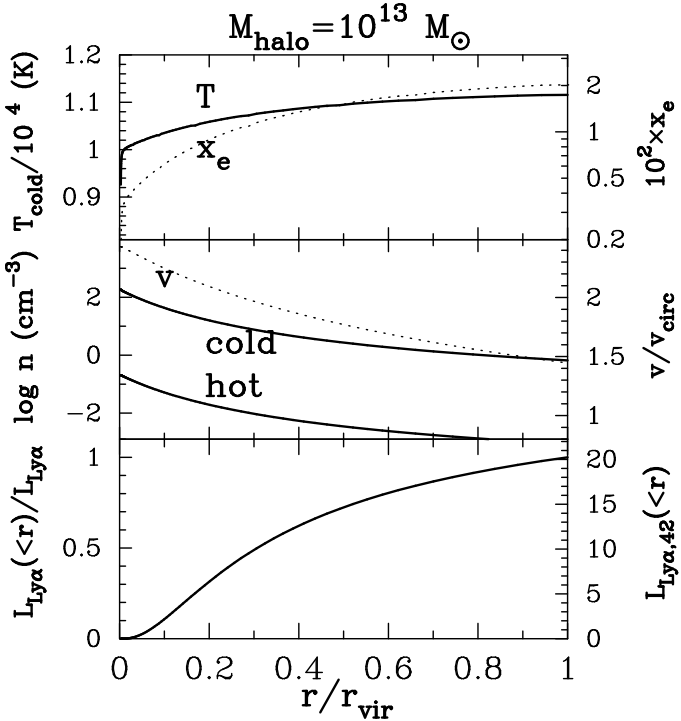}}}
\caption[]{Same as Fig~\ref{fig:profile}, but for a dark matter halo
of mass $M_{\rm halo}=10^{13} M_{\odot}$.}
\label{fig:profile2}
\end{figure}
Throughout the paper we assume that $f_{\rm grav}=0.3$, which we
regard as a conservative choice (see \S~\ref{sec:model} for a more
detailed discussion). Furthermore, we assume that
$\mathcal{T}_{\alpha}=0.5$. That is, $\sim 50\%$ of the emitted
Ly$\alpha$ photons is observed. The most likely source of opacity is
provided by residual HI in the intergalactic medium (IGM), which mostly affects the flux of photons that
was emitted blueward of the Ly$\alpha$ resonance \citep[e.g Fig~3
of][]{Madau95}, 
 At $z=2$, $z=3$ and $z=4$, the mean transmitted fraction of photons emitted blueward of the Ly$\alpha$ resonance is
$\langle\mathcal{T}_{\alpha}\rangle\sim0.8$,
$\langle\mathcal{T}_{\alpha}\rangle\sim 0.7$, and
$\langle\mathcal{T}_{\alpha}\rangle\sim 0.4$, respectively
\citep[e.g.][]{Claude2}. However, these fractions apply to an average
that was taken over a scales of tens of comoving Mpc (cMpc) along the
line of sight. Since the massive halos of interest reside in dense
parts of the Universe, we expect the local  to be more opaque than
average \citep[e.g.][]{IGM,D09}, which motivates our choice
$\mathcal{T}_{\alpha}=0.5$ at $z \leq 3$. The gas in the cold flows
consists mainly of compressed neutral intergalactic gas and dust
opacity is likely negligible (see \S~\ref{sec:dust} for additional
discussion on dust).

\subsection{The Halo Mass-Ly$\alpha$ Luminosity Correlation}
\label{sec:lvsm}

Consider a halo of mass $M_{\rm halo}=10^{12}M_{\odot}$, for which
$v_{\rm circ}=220$ km s$^{-1}$, $T_{\rm vir}=1.9 \times 10^6$ K, and
$r_{\rm vir}=82$ kpc (see, e.g. Barkana \& Loeb 2001, their
Eqs.~24-26).  The {\it solid line} in the {\it top panel} of
Figure~\ref{fig:profile} shows the radial dependence of the gas
temperature in the cold flow. The {\it dotted line} shows the electron
fraction by number, i.e. $x_e=x_{\rm HII}=n_{\rm HII}/n_{\rm p}$,
where $n_p$ denotes the number density of protons.  The temperature is
found to be in the narrow range of $\sim 1$--$1.1\times 10^4$ K which
results in a very low level of ionization ($x_e\sim 10^{-2}$). This is
caused by the fact that atomic line cooling is very efficient at the
densities encountered in the cold flow (see {\it central panel}), and
so the gas cools down to the temperature floor below which collisional
line excitation is ineffective. The temperature and ionization
fraction increase outward as the density, and therefore the cooling
rate decreases with radius. Figure~\ref{fig:profile2} shows the same
quantities for a halo of mass $M_{\rm halo}=10^{13}M_{\odot}$.

The {\it solid lines} in the {\it central panels} shows the number
densities of gas particles in both the cold and hot phase. The {\it
dotted line} shows the infall velocity in units of $v_{\rm circ}$.
(Note that the infall velocity enters the heating rate through
Eq.~\ref{eq:heat}.) The cold gas is denser by a factor of ${T_{\rm
vir}}/{T_{\rm c}}$. This results in number densities in the cold phase
that exceed $1-10$ cm$^{-3}$ at $r \lsim 0.5 r_{\rm vir}$. These high
densities justify our assumption of self-shielding: the recombination
rate in the cold flow is $\alpha_{\rm rec}n_{c}\sim 10^{-11}-10^{-12}$
s$^{-1}$, while the photoionization rate by the UV background in the
 IGM at $z=3$ is $\Gamma_{\rm UV} \approx 6 \times 10^{-13}$ s$^{-1}$
\citep[e.g][]{Claude}. Even the gas on the very edge of the cold flow,
that is optically thin to ionizing background, contains an ionized
fraction of hydrogen (by number) of only $x_{\rm
HI}=1+\frac{a}{2}-\frac{1}{2}(a^2+4a)^{1/2}$, where $a\equiv
{\Gamma_{\rm UV}}/{n_c\alpha_{\rm rec}}$. Therefore, $x_{\rm
HI}=0.2-0.6$ for $n_c=1-10$ cm$^{-3}$. That is, even this gas is
significantly neutral, and the gas becomes self-shielding less than a
pc into the flow. Note that boosting the local photoionization rate by
a factor of $10^2$ does not change this conclusion\footnote{Because in
our model the Ly$\alpha$ blobs are associated with massive halos,
$M_{\rm halo}>10^{12}M_{\odot}$, the cold filaments are compressed by
a factor of $(T_{\rm vir}/10^4\hs {\rm K})\gsim 200(M_{\rm
halo}/10^{12}M_{\odot})^{2/3}$ by the hot virialized gas. Therefore
the gas in our model is denser and better capable of self shielding
than in the models of \citet{Fur05}. For this reason, the
photoionizing background is not dominating the heating -- and
resultant Ly$\alpha$ cooling -- rate from the cold flows}.

The {\it lower panel} shows the total Ly$\alpha$ luminosity that was
emitted interior to radius $r$ (normalized to the total, which is
$L_{\alpha,{\rm tot}}\approx 2\times 10^{43}$ erg s$^{-1}$).  We find
that $\sim 50\%$ of all Ly$\alpha$ is emitted at $r\gsim 0.3r_{\rm
vir}$. Hence, our total calculated Ly$\alpha$ luminosity is emitted
over a spatially extended region. It is not possible to robustly
predict the surface brightness profile as this is determined by the
number of filaments in the cold flow, their geometry and their
orientation relative to the observer. The surface brightness profile
would furthermore be sensitive to any radial dependence of the
'gravitational heating efficiency'-parameter $f_{\rm grav}$.

\begin{figure}
\vbox{\centerline{\epsfig{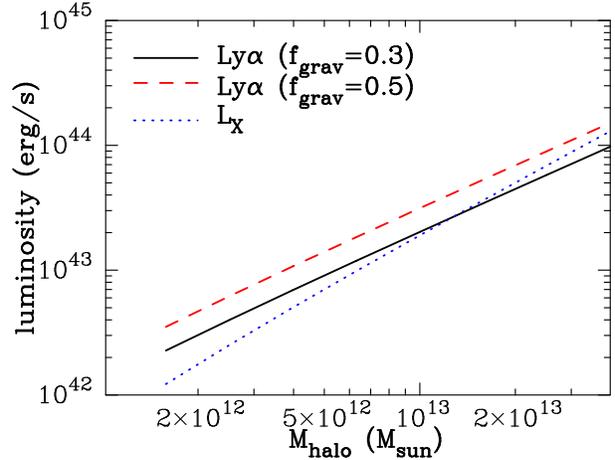}}}
\caption[]{The {\it solid line} shows the dependence of the total
Ly$\alpha$ luminosity as a function of halo mass, $M_{\rm halo}$, when
a fraction $f_{\rm grav}$ of the gravitational binding energy of the
cold gas is converted into heat. This model predicts Ly$\alpha$
luminosities in the range $10^{42}$--$10^{44}$ erg s$^{-1}$
(comparable to the range of values observed from the majority of
blobs, see Matsuda et al 2004) for halo masses in the range $M_{\rm
halo}=10^{12}-3\times 10^{13}M_{\odot}$. To explain the luminosity of
the most luminous ($L_{{\rm Ly}\alpha}=10^{44}$ erg s$^{-1}$) blob
with cold accretion requires a halo of mass $M_{\rm halo}\sim 3\times
10^{13}M_{\odot}$, close to the maximum halo mass that can host cold
flows \citep{Dekel08}. The {\it blue dotted line} shows the soft X-ray
(free-free) luminosity of the hot gas. There are not nearly enough
X-rays to significantly boost the Ly$\alpha$ emissivity of the cold
flows (see text).}
\label{fig:lvsm}
\end{figure}

Figure~\ref{fig:lvsm} shows the dependence of the total Ly$\alpha$
luminosity, $L_{{\rm Ly}\alpha}$, as a function of halo mass, $M_{\rm
halo}$ as the {\it solid line}. We find Ly$\alpha$ luminosities in the
range $10^{42}-10^{43}$ erg s$^{-1}$, comparable to the observed range
in LABs \citep{Matsuda04} for halo masses in the range $M_{\rm
halo}=10^{12}$--$10^{13}M_{\odot}$. In order to obtain $L_{{\rm
Ly}\alpha}=10^{44}$ erg s$^{-1}$, which is the luminosity of the
brightest of blobs that was found by \citet{Steidel00}, we either need
a halo with a mass of $\sim 3\times 10^{13}M_{\odot}$, or a  somewhat
lower halo mass with a higher $f_{\rm grav}$.

The {\it blue dotted line} shows the total X-ray luminosity of the hot
phase which is given by,
\begin{equation}
L_X=\int_0^{r_{\rm vir}}dr\hs 4\pi r^2 n_{\rm e}(r)n_{\rm
HII}(r)(1+Y_{\rm He})\epsilon_{\rm ff}(T_{\rm vir}),
\end{equation}
where $\epsilon_{\rm ff}(T_{\rm vir})=1.4\times 10^{-27}T^{1/2}$ erg
s$^{-1}$ cm$^{3}$ is the free-free emissivity \citep[see e.g. Eq~5.15b
of][]{RL79}. Note that the hot gas is fully ionized with $n_{\rm
HII}(r)=n_h(r)/(2+\frac{3}{4}Y_{\rm He})$, and an electron density
$n_e=(1+\frac{1}{2}Y_{\rm He})n_{\rm HII}$.

The X-ray luminosity is comparable to the Ly$\alpha$ luminosity. One
may think that the X-rays can penetrate the cold flows and boost the
Ly$\alpha$ luminosity. This is unlikely the case, as the frequency
dependence of the free-free spectrum is given by $L_{\nu,ff}\propto
\exp(-h_p\nu/kT_{\rm h})$, and therefore the hot gas emits photons as
energetic as $h_p\nu\sim 0.5(T_{\rm vir}/5\times 10^6\hs{\rm K})$
keV. These soft X-ray photons easily penetrate deep into the cold
flows where they photoionize either hydrogen or helium atoms, and
create an energetic electron. In a neutral medium, $\sim 40\%$ of the
energy of the electron goes into exciting HI, $\sim 30\%$ goes into
collisionally ionizing HI, and $\sim 10\%$ goes into heating the gas
(the rest excites and ionizes helium, Shull \& Van Steenberg
1985). Hence, even if the cold flows absorb all X-rays, line
excitation by secondary electrons would only boost the Ly$\alpha$
emissivity by at most a factor of $\lsim 1.4$. Furthermore, a boost in
the ionization (and thus electron) fraction naively boosts atomic line
cooling rate -- and hence Ly$\alpha$ emissivity -- by the same
factor. However, even if X-rays (significantly) enhance the electron
fraction in the cold flow, the cooling at a given temperature would
only be more efficient so as to balance the heating rate. Since at
most $10\%$ of the X-ray energy is converted into heat, this heating
rate is boosted by a factor $\lsim 1.1$. The above estimates assume
that the cold flows absorb {\it all} the X-rays; this is very unlikely
to occur in reality because the cold medium occupies a fraction $\sim
f_{\rm cold} T_{\rm c}/T_{\rm h}\sim 10^{-4}$--$10^{-3}$ of the volume
of the halo, and the majority of X-rays freely escape out from the
halo. The boost in Ly$\alpha$ emissivity from the cold flows due to
X-rays is therefore negligible. The X-ray photons that escape would
redshift to energies $h_p\nu\lsim 0.125 (T_{\rm vir}/5\times
10^6\hs{\rm K})$ keV, and would not be detectable with existing X-ray
telescopes. Note however, that metals can boost the X-ray emissivity
of the hot gas at energies $\gg 1$ keV. Deep X-ray observations may
therefore set useful limits on the amount of hot gas in the dark
matter halos hosting the blobs \citep[as in][who put an upper limit on
the average mass of halos hosting Ly$\alpha$ blobs of $M_{\rm halo}<
10^{13}M_{\odot}$]{Geach09}, although these limits do depend on the
assumed metallicity of the hot gas.

\subsection{The Ly$\alpha$ Luminosity Function of Cold-Accretion Powered Blobs}
\label{sec:lf}

The previously obtained relation between galaxy mass and Ly$\alpha$
luminosity allows us to compute a luminosity function of accretion
powered LABs from the mass function of dark matter halos. Observed
LABs are known to be associated with a region that contain a
significant overdensities of Lyman-break galaxies (LBGs) with a
fractional excess density of $\delta_{\rm LBG}=5.0\pm 1.2$
\citep[][]{Steidel98,Steidel03}. It is essential to take into account
this overdensity when predicting the expected number density of LABs
brighter than $L_{\alpha}$ as \citep{BL04},

\begin{equation} 
n_{\rm blob}(> L_{\alpha};\delta_{M}(V_{\rm s}))=\int_{\rm M_{\rm
min}(L_{\alpha})}^{\infty}dm\hs \frac{dn_{\rm
ST}}{dm}B(m,\delta_{M}(V_{\rm s})),
\label{eq:lf} 
\end{equation}
where $(dn_{\rm ST}/dm)$dm denotes the Sheth-Tormen number density of
dark matter halos (in cMpc$^{-3}$, Sheth \& Tormen 1999)  in the mass
range $M_{\rm halo}=m\pm dm/2$. The factor $B(m,\delta_{M}(V_{\rm
s}))$ denotes the boost in the number of halos due to the overall
matter overdensity, $\delta_{M}(V_{\rm s})$, within the survey volume
$V_{\rm s}$.

\begin{figure}
\vbox{\centerline{\epsfig{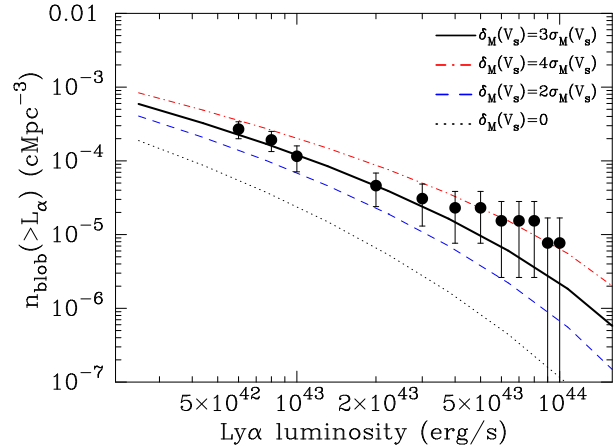}}}
\caption[]{The {\it solid line} shows the number density of Ly$\alpha$
blobs brighter than $L_{\alpha}$, $n_{\rm blob}(>L_{\alpha})$
cMpc$^{-3}$, and taking into account that fractional overdensity of
the region of the Universe in which the blobs reside is
$\delta_{M}(V_{\rm s})=3\sigma_M(V_{\rm s})$.  The {\it dashed} and
{\it dotted} line refer to values of $\delta_{M}(V_{\rm
s})=2\sigma_M(V_{\rm s})$ and $\delta_{M}(V_{\rm s})=5\sigma_M(V_{\rm
s})$, bracketing the likely range of overdensities inferred from the
observed overdensity of LBGs in the region by Steidel et al. (1998).
Here, $\sigma^2_M(V_{\rm s})$ denotes the variance of the density
field averaged over spheres of volume $V_s=1.3 \times 10^5$ cMpc$^{3}$
(the survey volume).  Our model with $f_{\rm grav}=0.3$ provides a
good fit to the data. The present uncertainty on $\delta_{M}(V_{\rm
s})$ allows models with $f_{\rm grav}\gsim 0.15$ to reproduce the
observed number of Ly$\alpha$ blobs. The {\it black dotted line} shows
the number of blobs expected in an 'average' part
(i.e. $\delta_{M}(V_{\rm s})=0$) of the Universe, indicating how rare
the most luminous blobs are ($n_{\rm blob}(>10^{44} {\rm erg}\hs{\rm s}^{-1})\sim 10^{-7}$ cMpc$^{-3}$).}
\label{fig:lf}
\end{figure}

In Figure~\ref{fig:lf} we plot $n_{\rm blob}(>
L_{\alpha};\delta_{M}(V_{\rm s}))$ as a function of Ly$\alpha$
luminosity for three values of the overdensity $\delta_{M}(V_{\rm
s})=2\sigma_M(V_{\rm s})$ ({\it dashed line}), $3\sigma_M(V_{\rm s})$
({\it solid line}), and $4\sigma_M(V_{\rm s})$ ({\it red dotted
line}). Here, $\sigma^2_M(V_{\rm s})=0.013$ denotes the variance in
the matter density field at $z=3$ averaged over spheres of the survey
comoving volume $V_s=1.3 \times 10^5$ cMpc$^{3}$. The overdensities
$\delta_M(V_s)$ were chosen so as to reproduce the observed boost and
standard deviation in the LBG number density, under the assumption
that LBGs are associated with dark matter halos of mass $M_{\rm
halo}=10^{12}M_{\odot}$ (see Appendix~\ref{app:delta}).
 
The data points were derived by taking the ratio between the number of
blobs \citep{Matsuda04} brighter than $L_{\alpha}$ and the survey
volume $V_{\rm s}=1.3 \times 10^5$ cMpc$^{3}$. The error bars on the
data points were obtained from the relation
$\sigma^2=b^2\sigma^2_M(V_{\rm s})+\sigma_{P}^2$, where $\sigma_P$
denotes the Poisson error and $b^2\sigma^2_M(V_{\rm s})$ denotes the
cosmic variance for a bias factor $b=b(M_{\rm halo},z)$, which is in
turn calculated following \citet{S04}. For the cumulative luminosity
function, the errors were added in quadrature.

Figure~\ref{fig:lf} shows that the model luminosity function agrees
well with the data \citep{Matsuda04}. The uncertainty in
$\delta_M(V_s)$ allows for a large spread in the predicted $n_{\rm
blob}(>L_{\alpha})$, especially at the bright end. This is because the
most massive ($M_{\rm halo}\sim 10^{13}M_{\odot}$) halos are highly
biased tracers of the mass and their number density depends
sensitively on $\delta_{M}(V_{\rm s})$. The sensitivity of the
cumulative luminosity function to $\delta_{M}(V_{\rm s})$ implies that
good fits to the data are possible for a range of gravitational
heating efficiencies, $f_{\rm grav}$. In the allowed range between
$\delta_{M}(V_{\rm s})=2\sigma_M(V_{\rm s})$ and $4\sigma_M(V_{\rm
s})$, we obtain a good fit for $f_{\rm grav} > 0.15$. Our predicted
number of Ly$\alpha$ blobs is practically degenerate in the product
${\mathcal T}_{\alpha}\times f_{\rm grav}$, and this constraint can be
recast as ${\mathcal T}_{\alpha} f_{\rm grav}\gsim 0.075$. Therefore,
only a relatively low gravitational heating efficiency suffices to
make the cold flows 'glow' in Ly$\alpha$ at luminosities observed from
the Ly$\alpha$ blobs.

Lastly, the {\it black dotted line} in Figure~\ref{fig:lf} shows the
number density of blobs expected in an 'average' part
(i.e. $\delta_{M}(V_{\rm s})=0$) of the Universe. This implies that
the most luminous blobs that were originally discovered by Steidel et
al (2000), are significantly rarer in the field (where $n_{\rm
blob}(>10^{44} {\rm erg}\hs{\rm s}^{-1})\sim 10^{-7}$
cMpc$^{-3}$). Our prediction is consistent with a recent upper limit
of $n_{\rm blob}(L_{\alpha}>10^{44}$ erg s$^{-1}) \lsim 5 \times
10^{-7}$ cMpc$^{-3}$ \citep{Yang08}.

\section{Discussion}
\label{sec:discuss}

\subsection{Predicted Ly$\alpha$ Blobs Spectral Line Shapes and Physical Sizes}
\label{sec:size}
The typical velocity widths (FWHM) of the LABs span the range
\citep{Matsuda06} of $500$--$1700$ km s$^{-1}$. In the cold-accretion
model, infall occurs at velocities in the range $1.0$--$2.5v_{\rm
circ}\sim 220$--$1500$ km s$^{-1}$, which naturally result in the
observed velocity widths (taking into account projection effects due
to the orientation of the filaments, and the intergalactic absorption
that suppresses the observed flux of the 'bluest' Ly$\alpha$ photons
\citep[e.g.][]{Madau95}). Although the gas in the cold flows is
mostly neutral, resonant scattering is not expected to broaden the
Ly$\alpha$ line beyond the observed velocity widths of the blobs, as
we discuss next.

The cold flows are highly elongated structures, and the majority of
the Ly$\alpha$ photons escape in a direction perpendicular to the axis
of the flow. Therefore, at the densities of interest the Ly$\alpha$
photons have to traverse an HI column density $N_{\rm HI}\sim
10^{21}-10^{22}(R_{\rm flow}/0.5\hs{\rm kpc})$ cm$^{-2}$ before
escaping from the flow. Here, $R_{\rm flow}$ denotes the radius
associated with the characteristic cross-sectional area of each
stream\footnote{The cross-sectional radius of a cold filament can be
estimated as follows. The total solid angle that is covered by the
cold gas at a distance $r$ from the center of the halo is $\Omega_{\rm
cold}(r)=4\pi \times(f_{\rm cold}\frac{T_c}{T_{\rm vir}})$. Cold
accretion typically occurs via multiple cold streams, and the average
solid angle that is covered by a single cold filament is $\Omega_{\rm
fil}(r)=\frac{1}{N_{\rm fil}}\times 4\pi \times(f_{\rm
cold}\frac{T_c}{T_{\rm vir}})$, where $N_{\rm fil}$ is the total
number of filaments. From this it follows that the angular size of a
single cold flow in radians, when viewed from the center of the halo,
is $\theta_{\rm fil}=\sqrt{\frac{1}{N_{\rm fil}}\times 4\times(f_{\rm
cold}\frac{T_c}{T_{\rm vir}}})$, which translates to a physical radius
of $R_{\rm flow}(r)=\theta_{\rm fil} r$. Substituting numbers yields
$R_{\rm flow}=0.8(N_{\rm fil}/5)^{-1/2}(r/0.5 r_{\rm vir})$ kpc (where
the mass dependence enters only through $f_{\rm cold}$ ).}. Scattering
through such large columns broadens the Ly$\alpha$ line to a width of
$\sim 800(N_{\rm HI}/10^{22}\hs{\rm cm}^{-2})^{1/3}(T_c/10^4\hs{\rm
K})^{1/6}$ km s$^{-1}$ \citep{N90}.
Velocity gradients that exist throughout the flow will reduce this
broadening: when velocity gradients are present, the relevant length
scale -- and hence column density -- that set the amount of velocity
broadening is the Sobolev length, defined as the length over which the
velocity changes by one thermal width $l_{\rm S}\equiv v_{\rm
th}/[dv(r,R)/dR]$ \citep[][and see e.g. \S~6.2 in Dijkstra \& Loeb
2008b]{Bo79}. Here, $R$ denotes the transverse direction in the cold
flow.  If for example, $dv/dR\sim v_{\rm circ}/R$ (which appears
reasonable for some cold filaments based on inspection of Figs.~8 \& 9
of Dekel et al 2009), then we get that the total amount of velocity
broadening is reduced by a factor of $(v_{\rm circ}/v_{\rm
th})^{1/3}\sim 2.6(M_{\rm halo}/10^{12}M_{\odot})^{1/9}$.  We
therefore conclude that although scattering plays a non-negligible
role in broadening the Ly$\alpha$ line profile, it does not lead to
Ly$\alpha$ velocity widths that exceed those of the observed
blobs. However, the complexity of the associated radiative transfer
effects does not allow us to make precise predictions for the
Ly$\alpha$ line width and its relation to the halo mass (or
corresponding Ly$\alpha$ luminosity).

Because of the low volume filling factor of the cold flows, we do not
expect the Ly$\alpha$ photons to scatter off neutral hydrogen (HI) in
separate cold filaments once they have escaped from their filament of
origin. For this reason, scattering will lead to a very low level of
polarization (or none at all) for these models, in difference from the
models of \citet{DL08} who predicted high levels of polarization for
Ly$\alpha$ emerging from neutral, spherically symmetric, collapsing
gas clouds. In addition, we do not expect the systematic blueshift of
the Ly$\alpha$ line that was predicted \citet{D06} for spherically
symmetric models\footnote{Interestingly, \citet{Adams08} did recently
detect a systematic blueshift of the Ly$\alpha$ line relative to the
observed 21-cm absorption line. They simultaneously reproduced both
observations with radiative transfer of Ly$\alpha$ through a
spherically symmetric, neutral collapsing gas cloud with a gas mass of
$M_{\rm gas}\sim 10^{12}M_{\odot}$. Unless the mass of the host dark
matter halo is implausibly high, $M_{\rm halo}\sim 10^{14}M_{\odot}$,
this would require an unusually high fraction of the gas to be in the
cold phase (with $f_{\rm cold}$ close to unity). This inference might
imply a significant scatter in $f_{\rm cold}$ for a given halo
mass. Some scatter in $f_{\rm cold}$ is indeed observed in
simulations, although not at the level to explain cold fractions close
to unity. The possible existence of large $f_{\rm cold}$ values would
only boost the predicted number of cooling powered blobs and
strengthen our basic conclusions.}.
 
Some resonant scattering would occur for photons that enter the IGM
 blueward of the line resonance. The impact of the IGM would be
frequency-dependent, and we expect the associated 'blurring' to be
most prominent for the most energetic Ly$\alpha$ photons. The
resonantly scattered Ly$\alpha$ radiation is expected to have a low
level ($\lsim 7\%$) of linear polarization \citep{DL08}.

The observed LABs have a large spatial extent \citep{Steidel00} of
$\sim 150$ kpc. In our simplified treatment, Ly$\alpha$ is emitted
from a region that extends out to the virial radius of the host halo,
\begin{equation}
r_{\rm vir}\approx 82 {\rm kpc} \left({M_{\rm halo}\over
10^{12}M_\odot} \right)^{1/3} \left({1+z \over 4}\right)^{-1}.
\end{equation}
This implies LABs with diameters of $\sim 160$--$350$ kpc for halo
masses in the range of $M_{\rm halo}\sim 10^{12}$--$10^{13}M_\odot$.
This range is expected to exceed the observed blob sizes, since our
estimate was derived under the simplifying assumption of a
gravitational heating efficiency $f_{\rm grav}$ that is independent of
radius. Higher resolution AMR simulations by \citet{Agertz09}
show that cold streams slow down progressively after they enter the
inner halo ($r\lsim 0.3-0.5r_{\rm vir}$), and therefore suggest that
$f_{\rm grav}$ is likely to increase at smaller radii. Models that
account for this increase predict smaller blob sizes that are closer
to the observed values. It is also possible that existing observations
only detect Ly$\alpha$ radiation from the inner regions of cold flows
where the Ly$\alpha$ surface brightness is high.  Indeed, $\sim
60-70\%$ of our computed Ly$\alpha$ luminosity emerges from
$r<0.5r_{\rm vir}$ (with this modest reduction factor having a
negligible effect on Fig.~\ref{fig:lf}).  In this case, deeper
observations would reveal filamentary extensions of the blobs to
larger scales.

\subsection{The Role of Dust}
\label{sec:dust}

Neutral gas in the cold flows forces each Ly$\alpha$ photon to scatter
multiple times before escaping from the flow. As a result, Ly$\alpha$
photons traverse on average a total distance through the cold flow
that is $\sim 130(N_{\rm HI}/10^{22}\hs{\rm
cm}^{-2})^{1/3}(T_c/10^4\hs{\rm K})^{1/6}$ times larger than that for
continuum photons \citep{Adams75}. One would therefore expect the
Ly$\alpha$ flux to be highly susceptible to absorption by dust. In
reality, dust extinction is probably not an important effect since the
cold flows consist of compressed intergalactic gas that has not been
processed through galaxies. We therefore expect these flows to contain
significantly less dust than the interstellar medium of star forming
galaxies.  Indeed, it is likely that quenching of Ly$\alpha$ flux by
dust is more important for superwind-generated emission (see
\S~\ref{sec:alt} for a discussion of alternative models for Ly$\alpha$
blobs), since in this case the Ly$\alpha$ flux is generated in cold
neutral shells of gas that were swept up from the interstellar medium
of the galaxy. \citet{Mori04} found the dust-opacity to Ly$\alpha$ of
their superwind generated shells to be negligible ($\tau_d\sim
0.03[N_{\rm HI}/5\times 10^{21}\hs{\rm cm}^{-2}]$). However, their
estimate of $\tau_d$ does not take into account the fact that
scattering can boost the total distance traversed by Ly$\alpha$
photons through the shell -- and therefore the optical depth--by a
factor of $\sim 110(T_c/10^4\hs{\rm K})^{1/6}$ to a value of
$\tau_d\sim 3.3(N_{\rm HI}/5\times 10^{21}\hs{\rm
cm}^{-2})(T_c/10^4\hs{\rm K})^{1/6}$. It is therefore a serious
concern that the dust opacity to Ly$\alpha$ photons in these models
exceeds unity.

On the other hand, the cold flows are expected to contain
significantly less dust because they originate in the , and so we
conclude that the Ly$\alpha$ flux generated in cold flows is not
subject to the same level of dust opacity. For a typical intergalactic
gas metallicity $Z_{\rm }\sim 10^{-3}Z_{\odot}$
\citep[e.g.][]{Schaye03}, and for typical HI column densities and
velocity gradients in the cold flows (see \S~\ref{sec:size}), we
expect a dust opacity that is well below unity. This low level of
enrichment is not unreasonable, as galactic outflows would typically
avoid the overdense inflowing filaments. Indeed, preliminary
simulations indicate that outflows generally enrich the large volume
surrounding the voids, while leaving the cold flows metal poor
(R. Joung, private communication).

\subsection{The `Duty Cycle' of the Cold Accretion Mode}
 
When computing the cumulative luminosity function of blobs through
Eq.~(2), we implicitly assumed a duty cycle of unity. That is, we
assumed each massive galaxy halo to be surrounded by cold
flows. Simulations support this assumption: the massive halos which
produce a detectable Ly$\alpha$ luminosity typically reside in
overdense regions of the Universe in which accretion of cold gas from
filaments onto the more massive halos is a continuous process at
redshifts $z\ga 2$. This is in sharp contrast to intense starburst
and/or quasar activity, which occur with duty cycles that are much
smaller than unity. Indeed, in our model the cold flows are luminous
in Ly$\alpha$ irrespective of the activity in the central galaxy,
which explains why only a fraction of Ly$\alpha$ blobs are observed to
be associated with intense starburst or quasar activities.

\subsection{Comparison to Other 'Blob Models'}
\label{sec:alt}

While the cold accretion model can successfully reproduce the
abundance of LABs in the Universe as well as their physical
properties, it is useful to compare our model to the other models for
LABs mentioned in \S~\ref{sec:intro}. The observed number density of
blobs, combined with their preferred residence in dense regions of our
Universe, requires an association with massive halos ($M_{\rm
halo}\sim 10^{12}$--$10^{13}M_{\odot}$).

An association with bright quasars also naturally places the blobs in
massive halos \citep{H07}. However, the filamentary geometry of the
simulated cold gas shield it from any central source of ionizing
radiation, suppressing the photoionization rate by the
quasar. Furthermore, the short duty cycle of quasar activity
\citep{M04} is problematic. Indeed, a large fraction of blobs have no
associated quasar nearby \citep[see][for examples of Ly$\alpha$ blobs
associated with radio quiet type I Active Galactic Nuclei
(AGN)]{Bunker03,Weidinger04}.

The duty cycle argument, combined with the lack of the relevant
associated sources, argues against photoionization by inverse Compton
radiation, or star formation that is triggered by AGN jets. 
Formally, photoionization by regular star forming galaxies (LBGs) is
not ruled out for $\sim 60\%$ of Ly$\alpha$ blobs
\citep[e.g.][]{Matsuda04}. However, it remains to be explained why the
LBGs in these dense environments would be so efficient at generating
Ly$\alpha$ radiation, while observationally it is known that the
escape fraction of Ly$\alpha$ radiation from LBGs is typically low,
and that only $20-25\%$ of $z=3$ LBGs would even classify as a
Ly$\alpha$ emitting galaxy \citep{Shapley03}. Indeed, for this reason 
\citet{JH06} argued for photoionization by primordial galaxies which
can emit more ionizing photons per observed rest-frame UV
continuum. This is an interesting suggestion, although it remains to
be shown that the gas in highly overdense, evolved, regions of our
Universe can remain pristine down to $z=3$ at a level that is needed
to explain the Ly$\alpha$ blobs.
 
In any case, all photoionization models require cold spatially
extended gas to be present in these halos, because the gas at the
virial temperature of these halos ($T_{\rm vir} \gsim 2\times 10^6$ K)
would recombine too slowly to reproduce the observed Ly$\alpha$
luminosities.  Indeed, \citet{HR01} obtain their Ly$\alpha$
recombination radiation -- following photoionization by the central
quasar -- from cold ($T=10^4$ K) gas clouds embedded in a hot medium,
very similar to the cold-accretion scenario that is seen in
galaxy-formation simulations (the main difference being that the cold
phase consists of separate clouds in the early models [such as in Fall
\& Rees 1985] rather than long continuous streams). Because
photoionization models require the existence of cold spatially
extended gas, they may be viewed as a special case of the cold
accretion model in which the heating rate $H(r)$ is dominated by
photoheating (although it remains to be shown that the gas does not
self-shield). In principle the photoionization models also work when
the cold gas is supplied by outflowing material (see below). However,
dust in outflow models is a bigger problem here than for the inflow
models (see \S~\ref{sec:dust}).
 
A fundamentally different model for blobs is the superwind model. The
association of some blobs with sub-mm sources imply a connection with
starburst galaxies (having star formation rates of $\sim
10^{2-3}M_{\odot}$ yr$^{-1}$). The kinetic energy associated with the
starburst-driven superwinds is sufficient to power the Ly$\alpha$
emission in LABs \citep{Taniguchi00,Geach05}.  However, it is not
clear that this coupling of energy occurs naturally at the level that
is required to explain the LABs. \citet{Mori04} used hydrodynamical
simulations to show that superwinds driven by a starburst
($\dot{M}_{*,{\rm max}}\sim 200 M_{\odot}$ yr$^{-1}$) can produce an
LAB that closely resembles one of \citet{Steidel00} blobs in
appearance, but with a luminosity that is $\sim 10/{\mathcal
T}_{\alpha}$ times too small (note that ${\mathcal
T}_{\alpha}$$<1$). If we simply scale up the luminosity of an LAB in
proportion to the total star formation rate, then the observed
Ly$\alpha$ luminosity requires a star formation rate of $\sim
2000/{\mathcal T}_{\alpha}$ $M_{\odot}$ yr$^{-1}$. However, this is
close to the observed star formation rate derived from the detected
sub-millimeter flux in only one of Steidel et al.'s
LABs\citep{Geach05}. Also, the conversion from star formation rate to
the blob's Ly$\alpha$ luminosity assumed that the dust opacity for
Ly$\alpha$ photons is negligible. We showed in \S~\ref{sec:dust} that
this may not be correct when radiative transfer effects are accounted
for. Furthermore, not all blobs have starburst activity associated
with them \citep{Matsuda06,Nilsson06,Smith07,Saito08,Smith08b}, thus
ruling out this model for these sources.
 
\subsection{Testable Predictions of the Cold Accretion Model For Ly$\alpha$ Blobs}
\label{sec:obs}

Simulations suggests that the cold flows enter the center of the halo
as multiple filamentary streams. This filamentary structure is
expected to show up in the observations. In \S~\ref{sec:size} we
argued that some 'blurring' of the Ly$\alpha$ image may be expected,
but only for the most energetic Ly$\alpha$ photons, as these are
expected to resonantly scatter in the IGM. Existing images of LABs
typically have an angular resolution of $\sim 1$ arcsec, which
corresponds to $\sim 8$ kpc at $z=3$. Therefore, existing observations
would not resolve the filaments yet, although the filamentary nature
of the cold flows may show up at larger radii (where different
filaments are well-separated). Existing observations of blobs that are
presently thought to be powered by cold accretion do show some
interesting irregularities in their images \citep{Nilsson06,Smith08a},
but the quality of the data does not allow to classify the image as
'filamentary'. It is intriguing that the images of some of the LABs
discovered by \citet{Matsuda04} are very irregular, and possibly
filamentary (e.g. blob 6, 9, 11, and 12). If these blobs are
associated with cold flows, then we expect deeper and higher
resolution images of the blobs to more clearly reveal the filamentary
structure.
 
We argued in \S~\ref{sec:size} that we do not expect the systematic
blueshift of the Ly$\alpha$ line, nor the high level of polarization,
that were predicted by spherically symmetric models of neutral
collapsing gas cloud. This is mostly because the Ly$\alpha$ photons
scatter mostly in the filament where they were produced, after which
they escape to the observer (following some resonant scattering in the
IGM). Because the fraction of cold accreting gas increases towards
lower galaxy masses (and therefore lower blob luminosities in our
model), we may expect cold accretion to be less filamentary at lower
galaxy masses. This implies that the predictions of the models that
assumed spherically symmetric accretion -- involving the blueshift and
the high levels of linear polarization -- may be increasingly relevant
at lower halo masses and blob luminosities. However, these trends are
rather weak in the limited range of halo masses of interest here,
$M_{\rm halo}=10^{12}$--$10^{13} M_{\odot}$ (see e.g. Fig~5 and Fig~6
of Keres et al. 2009, but see Adams et al. 2009 and \S~\ref{sec:size}).
 
In our model, the vast majority of the Ly$\alpha$ radiation is
generated from collisional excitation of atomic hydrogen. At
temperatures of $T_{\rm c}\sim 10^4$ K, collisional excitation of the
Ly$\beta$ transition is $\sim 10$ times less effective. Therefore, we
expect an H$\alpha$ ($\lambda=6536$ \AA) flux that is $\sim
10{\mathcal T}_{\alpha}\times 6536/1216\sim 54{\mathcal T}_{\alpha}$
times lower. This is unfortunate because a combined measurement of the
Ly$\alpha$ and H$\alpha$ fluxes could have constrained the importance
of radiative transfer effects. In \S~\ref{sec:size} we concluded that
radiative transfer effects were not negligible, and that they may
noticeably broaden the Ly$\alpha$ line. To find this broadening
requires the detection of the much fainter H$\alpha$ line. It is not
guaranteed that the H$\alpha$ flux will actually be suppressed by a
factor as large as that mentioned above. As we argued in
\S~\ref{sec:unc}, the weak shocks that heat the cold flows on their
way to the halo center, may occasionally heat the gas to a temperature
at which hydrogen becomes significantly ionized. In this case, the
H$\alpha$ flux may be down by the value expected from recombinations
($\sim 9{\mathcal T}_{\alpha}$). These shocks may also result in
helium becoming singly ionized, leading to a detectable He 1640
emission \citep{Yang06}.
 
The observed LAB luminosity function (which we compiled from the data
presented by \citet{Matsuda04}) is reproduced by models in which
${\mathcal T}_{\alpha}$$f_{\rm grav}\sim 0.15$ (for $\delta_{M}(V_{\rm
s})=3\sigma_M(V_{\rm s})$, see Fig.~1).  Assuming that ${\mathcal
T}_{\alpha}$$f_{\rm grav}\sim 0.15$ (independent of redshift), we can
predict Ly$\alpha$ luminosity functions for accretion powered LABs at
$z=2$ and $z=4$ (when averaged over large regions of our Universe,
such that $\delta_{M}(V_{\rm s})\sim 0$). These luminosity functions
are shown in Figure~5, which implies that a decline in the number
density of accretion powered blobs more luminous than a few $\times
10^{43}$ erg s$^{-1}$ is expected beyond $z=3$. Such a decline -- if
present -- could be detected with future observations. For example,
the {\it Hobby-Eberly Telescope Dark Energy Experiment}
\citep[HETDEX,][]{Hill08} is expected to detect 0.8 million Ly$\alpha$
emitting galaxies and therefore thousands of LABs at $1.9<z<3.5$. Such
a large sample of blobs will allow for accurate measurements of their
spatial clustering, and luminosity functions. The combination of these
measurements should place tight constraints on blob formation models.

\begin{figure}
\vbox{\centerline{\epsfig{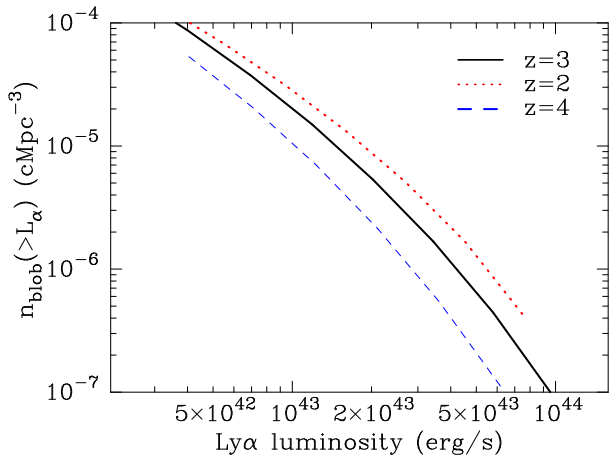}}}
\caption[]{The predicted cumulative luminosity functions of
cold-accretion powered LABs at $z=2$ ({\it red dotted line}), $z=3$
({\it solid line}), and $z=4$ ({\it blue-dashed line}) (in a region at
mean cosmic density) for $f_{\rm grav}{\mathcal T}_{\alpha}$$=0.15$
(the value preferred by fitting to the data of Matsuda et al 2004; see
text). A drop in the number density of sources more luminous than a
few $\times 10^{43}$ erg s$^{-1}$ is expected beyond $z=3$. This drop
may be enhanced by the  that becomes increasingly opaque with
increasing redshift.}
\label{fig:lf2}
\end{figure}

\subsection{Discussion of Model Uncertainties}
\label{sec:unc}

We have taken the total fraction of the gas that is in cold streams 
as a function of halo mass from the most recent simulations of cold streams
 in massive halos (see \S~\ref{sec:intro}). Other model assumptions that were described in \S~\ref{sec:model} were based on these same simulations. However,
the cold flow properties in our model - in particular the gas density
and temperature - can be quite different from what is seen in the
simulations. These differences are mostly due to the simplified nature
of our model which assumes a two-temperature baryonic component, while
in reality there is a wider spread in temperatures in the cold streams
(a more detailed discussion follows below). We have not attempted to
bring our model in closer agreement to the simulations because it
would require more detailed modeling of increasingly uncertain
physical processes. Most importantly, we will show below that these
differences are not important as far the Ly$\alpha$ emissivity of the
cold gas is concerned. In the following we address the differences in
cold flow properties between our model and the simulations, and
discuss other model uncertainties in more detail.

\subsubsection{Our Model versus Simulations: Cold Flow Temperature and Density}
\label{sec:T}

Our models predict gas densities that are significantly higher than those
observed in AMR simulations \citep[e.g.][]{Dekel08}. For a $M_{\rm
halo}=10^{12}M_{\odot}$ halo, we find $n_c \sim 1 $ cm$^{-2}$ at
$r=0.3r_{\rm vir}$, which is about a factor of $\sim 30$  larger than
the mean cold flow density that is found in the simulations. Similarly, the hot gas in our model denser
by about a factor of $\sim 3-4$ (A. Dekel, private
communication). There are several reasons for this discrepancy: ({\it
i}) the simulated flows are at a somewhat lower redshift ($z=2.5$)
than we consider ($z\sim 3$), which reduces the density by a factor of
$(4/3.5)^3\sim 1.5$; ({\it ii}) the cold fraction of gas is slightly
higher in AMR simulations ($f_{\rm cold,AMR}=0.4$--$0.5$ at $M_{\rm
halo}=10^{12}M_{\odot}$, see Birnboim et al. 2007) than in the SPH
simulations, which reduces the number density of particles in the hot
phase. This amounts to an additional factor of $(1-f_{\rm
cold,SPH})/(1-f_{\rm cold,AMR})\sim 1.5$. When combined, this explains
a factor of $(1.5)^2=2.25$. The remaining difference in the hot
gas density is at the factor
$\lsim 2$ level, which may be due to the details of gas density
profile that was used (also see \S~\ref{sec:hot}).

The fact that the density discrepancy is stronger for the cold gas suggests
that pressure equilibrium is not exactly satisfied in the simulation.
Indeed, in AMR simulations $P_{\rm cold}\sim 0.5 P_{\rm hot}$ at $r
\lsim 0.5r_{\rm vir}$ (A. Dekel, private communication) which implies
that at a given temperature, the number density in the simulated cold gas is
reduced by an extra factor of $\sim 2$. Furthermore, the gas temperatures
 in the simulated cold gas are somewhat higher
 (a more detailed discussion of this follows below)
which can lower the density by an extra factor of $\sim
2$. When collecting all factors of $2$ and $1.5$, an order of
magnitude difference in gas density can be accounted for. The remaining difference in the cold gas density is at the factor $\sim 3$ level, which may again be due to the details of gas density
profile that was used (also see \S~\ref{sec:hot}).

We emphasis that because of our requirement that the 
cooling rate balances the local heating rate, $H(r)$, 
{\it our predicted Ly$\alpha$ luminosity is not
affected by the exact density in the cold gas at all}. 
This is because the heating rate (Eq~\ref{eq:heat}) is independent of
gas density and temperature. Furthermore, radiative cooling is dominated by the emission of
Ly$\alpha$ photons at the typical cold gas
temperatures. We therefore indirectly regulate the Ly$\alpha$
emission rate {\it per hydrogen atom} by the local heating
term. When the cold flow gas density is  lowered/raised, a fixed cooling rate
is maintained by raising/lowering the cold gas temperature. For a fixed cold gas mass, we
therefore predict a fixed Ly$\alpha$ luminosity that depends almost entirely on
the heating rate. Formally, there {\it is} a (very) weak temperature
-and hence density- dependence of what fraction of the total cooling
(and therefore heating) rate emerges as Ly$\alpha$ line photons. For
example, at the temperatures that we encounter in our model $\sim
63-67\%$ of the all cooling radiation is in Ly$\alpha$ line photons,
while at T=$2\times 10^4$ K this fraction reduces to $\sim 50\%$. The
resulting very weak density dependence of the predicted Ly$\alpha$ 
luminosity has no impact on our results.

In our model, the halo gas consists of 'hot' and 'cold'
components. The hot gas is at virial temperature of the halo, $T_{\rm
vir}=1.9\times 10^6 {\rm K} (M_{\rm halo}/10^{12} M_\odot)^{2/3}
[(1+z)/4]$, while the cold flows have a typical temperature of $T_{\rm
c}\sim 0.95$--$1.2\times10^4$ K due to efficient cooling (as the
atomic cooling rate of the gas is high above $10^4$K and drops sharply
below this value). Simulations show a wider spread in temperatures of
the cold gas, namely $T_{\rm c}\sim 10^4$--$10^5$ K.  This apparent
discrepancy is not serious, as the gas at $10^5$ K (in the SPH
simulations) is mostly concentrated near the outermost regions of the
halo where cooling is less efficient. The vast majority of the gas in
the inner regions is in the range $T\sim 1-2\times 10^4$
K. Furthermore, the SPH simulations do not take into account
self-shielding of the gas (D. Keres, private communication), which
implies that photoheating is boosted artificially. More importantly,
ignoring self-shielding artificially suppresses the efficiency with
which the gas can cool (e.g. see Fig~2 of Katz et al. 1996). This at
least partially explains the somewhat higher temperatures observed in
these simulations.

Lastly, the processes that are responsible for heating the cold flows
may occasionally heat the gas to higher temperatures, $T \sim 2\times
10^4$--$10^5 $ K, after which it would rapidly cool to lower
temperatures. The warmest regions could be bright in emission of the
He 1640 \AA\hs line \citep{Yang06}, but at levels that are at most
$\sim 1$--$2$ orders of magnitude lower than the total flux in the HI
Ly$\alpha$ line (which is barely affected by these rare warmer regions
inside the cold flow). Hydrogen is mostly ionized at $T\sim 2\times
10^4$ K, which results in recombination radiation. This would boost
the expected H$\alpha$ flux emitted by the cold flows relative to the
H$\alpha$-flux expected from purely collisionally excited HI (see
\S~\ref{sec:obs}).

An important caveat to the simulations of cold flows is that they do
not incorporate superwinds that are generated by star-forming
regions. Superwinds can drive large-scale outflows that act as a
source of Ly$\alpha$ emission \citep{Mori04}. These winds expand most
efficiently into the lower density regions and will probably not
affect the cold flows because of the small solid angle occupied by the
cold filaments. Indeed, preliminary analysis of simulations that do
take into account winds support this claim (D. Keres, private
communication). Therefore, winds are expected to provide an additional
source of Ly$\alpha$ radiation, but would not disrupt the formation of
Ly$\alpha$ photons by gravitational heating of cold streams. Below we
show that this additional 'wind-generated' Ly$\alpha$ flux is likely
comparable or less than the flux from the cold flows.

\subsubsection{Energy Dissipation in the Hot Gas and Dynamical Instabilities in the Cold Flows} 
\label{sec:hot}

We assumed that no energy was dissipated in the hot gas, i.e. $f_{\rm
hot}=0$ in Eq~\ref{eq:vel}. This assumption was motivated by the
simulations in which the majority of the cold gas mass is in diffuse
continuous streams. This implies that the majority of cold gas is
surrounded by other cold gas, and that therefore little
interaction occurs between the hot and the majority of the cold gas. However,
shearing motions at the interface of the cold and hot
 gas may trigger the formation of instabilities that are not captured
by existing simulations of cold flows. In particular, Kelvin-Helmholtz
instabilities that develop at the cold-hot gas interface could
completely disrupt the cold
stream\footnote{http://mngrid.ucsd.edu/~akritsuk/renzo/kelvin/kelvin.html},
but this requires the relative velocity of the cold and hot gas to be
subsonic in both frames \citep[e.g.][and references
therein]{Miles57,BW95}. Because the sound speed in the cold gas is
$c_{\rm s,{\rm cold}}\sim 13$ km s$^{-1}$, this requirement is clearly
not met. However, the formation of instabilities in cold streams is
clearly an issue that needs to be addressed in future work. Instead of
disrupting the cold streams, these instabilities can introduce
non-negligible amounts of energy dissipation in the hot
gas\footnote{Interestingly, these same instabilities would heat the
cold gas through weak shocks, thus providing another physical
mechanism by which gravitational binding energy is converted into heat
in the cold gas.}.

\begin{figure}
\vbox{\centerline{\epsfig{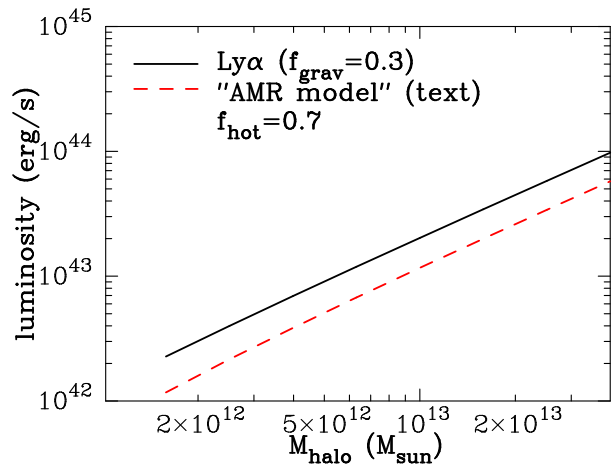}}}
\caption[]{The same as Fig~\ref{fig:lvsm}. Here, the {\it red dashed
line} represent a model in which a significant fraction of the binding
energy is dissipated in the hot gas, $f_{\rm hot}=0.7$. Furthermore,
we modified densities in the hot and cold gas to more closely resemble
those seen in AMR simulations (see text). The overall predicted
Ly$\alpha$ luminosities are reduced by a factor of $\sim 2.$. This can
be compensated for by increasing $f_{\rm grav}$ to $f_{\rm grav}=0.5$.}
\label{fig:fhot}
\end{figure}

We investigate the effect of energy dissipation in the hot gas in
Figure~\ref{fig:fhot}. Here, we plot Ly$\alpha$ luminosity as a
function of halo mass for a model a model in which $f_{\rm hot}=0.7$,
i.e. $\sim 70\%$ of the gravitational work done on an H-atom goes into
heating the hot gas, instead of going into bulk infall motion of the
atom (Eq~\ref{eq:vel}). Furthermore, for completeness we adjust
the cold flow properties to more closely resemble those encountered in
AMR simulations. In this 'AMR model' we({\it i}) assume different
hot gas density and temperature profiles as described in \S~2.1 of
Dekel \& Birnboim (2008, their model with $\alpha_g=0$), which appear
to provide a better description of the hot gas component in the AMR
simulation, ({\it ii}) reduce the overall gas density by reducing
the total gas mass fraction to $f_{\rm gas}=0.05$, ({\it iii}) boost the overall fraction of the gas in the cold phase by a factor
of $2$, ({\it iv}) assume that $P_{\rm cold}=0.5 P_{\rm
hot}$. Modifications ({\it ii-iv}) reduce the gas density in both the
cold and hot phase, and slightly increase the gas temperature in the
cold gas (but not to the values that are encountered in the
simulations, see \S~\ref{sec:T}). The overall reduction in the amount
of gas by a factor of $\sim 0.08/0.05=\sim 1.6$ is compensated for by
the boost in the cold fraction. However, the reduced infall velocity
reduces the local heating rate and thus the overall Ly$\alpha$
luminosity for a given halo. In order to restore the original halo
mass- Ly$\alpha$ luminosity relation, we require that $f_{\rm
grav}=0.5$. Note that it is possible that $f_{\rm grav}+f_{\rm hot}>1$, and it
corresponds to a model in which the cold gas decellerates as it
descends down the gravitational potential well. In order for solutions
to exist for $v(r)$ at all radii, we require that $f_{\rm grav}+f_{\rm
hot} \lsim 1.35$. Therefore a significant of the gravitational binding
energy may be dissipated in the hot gas without invalidating our
model, and for the example discussed above $f_{\rm hot}\lsim 0.85$ is
tolerated.

\subsection{Comparison to Previous Work} 

The high density gas in the cold flows can cool very efficiently, and
without any heat source this gas would almost instantly cool down to
temperatures at which the Ly$\alpha$ emissivity would be practically
zero. Because the gas densities in the cold flows are high, this gas
is capable of self-shielding and photoheating is negligible. We
therefore resorted to 'gravitational heating', which refers to the
hydrodynamical heating that the cold gas undergoes as it navigates
down to the center of the dark matter potential. This gravitational
heating mechanism was described in an analytic model by \citet{HS00}
who predicted the existence of Ly$\alpha$ blobs as a result of this
cooling radiation. \citet{Fardal01} obtained similar results using
hydrodynamical simulations -which already contained cold streams of
gas, albeit at significantly lower resolution. In both papers, the gas
emitted all its gravitational binding energy as cooling radiation
prior to reaching the central galaxy. This scenario would translate to
$f_{\rm grav}\geq1$ in our model.
Our work may be viewed as an important improvement over - or an
extension of - previous work, because our model is likely immune to
various important objections that can be raised against older models:
({\it i}) we have shown that a significant of the gravitational
binding energy may be dissipated in the hot gas without invalidating
our conclusions (\S~\ref{sec:hot}), ({\it ii}) in our model, gas
accretion occurs along dense, compressed, flows that cover a tiny
solid angle when viewed from the center of their host halo. Therefore,
these cold flows could not be easily disrupted by powerful outflows
from starburst activity or AGN, which would preferentially propagate
into the lower density (volume filling) phase of the halo gas. This
implies that cold flows can coexist with powerful sources, as observed
for some of the LABs \citep{Chapman01,Basu04,Geach07,Geach09}. For the
same reason, we do expect these outflows to enrich the inflowing cold
streams with metals and dust, which could severely quench the
Ly$\alpha$ flux from the cold streams (see \S~\ref{sec:dust}).

More recently, \citet{Fur05} used high resolution SPH simulations to
study Ly$\alpha$ emission from structure formation, which included
cold flows. In their highest resolution simulations, \citet{Fur05}
have a spatial resolution of $\sim 1$ kpc which is comparable to that
of \citet{Keres08}, but in a cosmological volume that is too small to
contain the massive halos that we associate with Ly$\alpha$
blobs. Despite the absence of these massive halos, these simulations
contained luminous blobs with properties that resemble the observed
ones. Hence, these models -just like the models of \citet{HS00} and
\citet{Fardal01}- overproduce the number density of blobs: existing observations
suggest that the field SSA21 contains a number
density of blobs that is $\sim 10-10^2$ larger than the Universe as a
whole (see \S~\ref{sec:lf} and Fig~\ref{fig:lf}). Ly$\alpha$ blobs
are likely rarer than was originally thought.

Simulations thus seemed to create luminous Ly$\alpha$ blobs ($L_{{\rm
Ly}\alpha}\sim 10^{44}$ erg s$^{-1}$) around too abundant lower mass
halos. This may partly be due to the fact that simulations contain a
non-negligible fraction of cold gas that is locked up into denser clumps
which reside both inside and outside the cold flows. Because of their
enhanced density, most of the Ly$\alpha$ luminosity would come from
these discrete clumps. These clumps may also be sites in which stars
form which may boost the Ly$\alpha$ emissivity from these clumped
regions even more. However, it is likely that Ly$\alpha$ emission does
not escape efficiently from star forming regions at $z=3$:
observations of $z=3$ Lyman break galaxies (LBGs) suggest that $\sim
20-25\%$ of star forming galaxies has a strong enough observed
Ly$\alpha$ emission line to classify as a Ly$\alpha$ emitting galaxy
\citep{Shapley03}. Furthermore, a significant fraction of these
Ly$\alpha$ emitting galaxies have emission lines that are weaker than
expected on the basis of recombination theory (Dijkstra \& Westra
2009). Without a good model for how Ly$\alpha$ escapes from these
clumps, their Ly$\alpha$ luminosity may well have been 
overestimated. This could explain why simulations have overpredicted
the number density of Ly$\alpha$ emitting blobs. Alternatively, the
simulated cold flows are too warm \citep[see e.g.][for a discussion on
why gas temperatures in self-shielded regions are notoriously
difficult to simulate]{Fur05}. As was mentioned earlier
(\S~\ref{sec:hot}), the gas temperature in our 'AMR model' was below that encountered in the simulation. This suggests that the AMR simulation also associates (much) more luminous
Ly$\alpha$ blobs to halos of a given halo mass than our model. This may result in a predicted number density of Ly$\alpha$ blobs that is at odds with the observations.

\section{Conclusions}
\label{sec:conc}

Recent hydrodynamical simulations of the formation of galaxies show
that baryons assemble into galaxies through a two-phase medium which
contains filamentary streams of cold ($T\sim 10^4$ K) gas in pressure
equilibrium with a hot gaseous halo
\citep[e.g][]{Keres05,Ocvirk08,Dekel08,Keres08,Agertz09}. These cold
flows contain $\sim 5$--$25\%$ of the total gas content
\citep{Keres08} in halos as massive as $M_{\rm halo} \sim$ a few
$10^{13}M_{\odot}$ \citep{Dekel08}.

At the typical densities and scales of the cold flows ($n_c\gsim 1$
cm$^{-3}$), the gas is self-shielded from external ionizing radiation
(see \S~\ref{sec:lvsm}) and is heated through gravitational
contraction. We have demonstrated that if $\gsim 10\%$ of the change
in the gravitational binding energy of a cold flow goes into heating
of the gas, then the simulated cold flows are spatially extended
Ly$\alpha$ sources with luminosities and number densities that are
comparable to those of observed Ly$\alpha$ blobs (see
Fig~\ref{fig:lf}).

Furthermore, the typical velocity widths of the LABs span the range
\citep{Matsuda06} of $500$--$1700$ km s$^{-1}$, which is consistent
with the model in which infall occurs at velocities in the range
$1.0$--$2.5v_{\rm circ}\sim 220$--$1500$ km s$^{-1}$, which naturally
result in the observed velocity widths (see \S~\ref{sec:size}. The
filamentary structure of the cold flows may explain the wide range of
observed Ly$\alpha$ blob morphologies.

The association with massive halos naturally places LABs in overdense
regions \citep{Steidel00,Matsuda04,Matsuda06}. Furthermore, the
simulated cold flows are dense and cover only a small solid angle when
viewed from the center of their host halo. Therefore, they could not
be easily disrupted by powerful outflows from starburst activity or
AGN, which would preferentially propagate into the lower density
(volume filling) phase of the halo gas. This implies that cold flows
can coexist with powerful sources, as observed for some of the LABs
\citep{Chapman01,Basu04,Geach07,Geach09}. The association of LABs with
sub-mm sources or AGN is not surprising, since these sources are
triggered by the infall of cold gas into massive galaxies. Even when
the central sources are sufficiently energetic to power the observed
Ly$\alpha$ emission, this causal relationship may be difficult to
achieve (see \S~\ref{sec:alt}). Our model associates the spatially
extended Ly$\alpha$ emission with the independent process of cooling
within the dense streams of inflowing cold gas. This resolves the
puzzling observation that very similar LABs may have completely
different sources associated with them \citep{Geach07,Yang08}.

Of course, it is possible that cold flows are partially photoionized
by an associated hard X-ray source \citep[as is observed for $\sim
17\%$ of the blobs][]{Geach09}, or by star forming galaxies that are
embedded within the cold flow \citep{Fur05}. In these cases, the
heating rate would be (locally) boosted. This would only make our
total predicted Ly$\alpha$ luminosities from the cold flows larger,
and would strengthen our conclusion that Ly$\alpha$ blobs are an
observational signature of cold accretion into galaxies.

Regardless of the precise heating mechanism, the cold flows should
reveal their filamentary geometry in deeper and or higher resolution
images (see \S~\ref{sec:obs}). Once this gas geometry has been probed,
it may be possible to constrain the dominant heating mechanisms of the
cold gas as it navigates toward the center of the dark matter halo
potential. Alternatively, if one does not find any evidence for
filamentary spatially extended Ly$\alpha$ sources in the near future,
then this is equally interesting, because this may suggest that the 
simulated cold flows are different than the flows that occur in nature, 
or - worse - do not represent reality at all (see \S~\ref{sec:hot}
 for a discussion on instabilities that may affect cold flow properties). 
In either case, Ly$\alpha$ observations are expected to provide us with 
unique constraints on this intriguing mode of gas accretion onto galaxies. 

{\bf Acknowledgments} This work is supported by Harvard University
funds. We thank Du\v{s}an Kere\v{s}, Yuval Birnboim, Zolt\'{a}n
Haiman, Ryan Joung, Renyue Cen, Avishai Dekel, Claude-Andr\'{e}
Faucher-Gigu\`{e}re, Matt McQuinn and Adam Lidz for useful discussions.

\label{lastpage}

\appendix
\section{The Ly$\alpha$ Blob Luminosity Function in an Overdense Region}
\label{app:delta}
\citet{Steidel98} found an overdensity of $\delta_{\rm LBG}=5.0\pm
1.2$ Lyman Break Galaxies \citep[LBGs][]{Steidel03} per unit volume
within their survey volume of $V_{\rm s}\sim 0.15 \times 10^5$
cMpc$^{3}$. This highly overdense region is contained within the approximately nine times larger volume, $V_{\rm S}$ (the subscripts 's' and 'S' refer to the smaller and larger survey volumes, respectively), in which the Ly$\alpha$ blobs were found
\citep{Matsuda04}. In later work, \citet{Matsuda05} found that the
Ly$\alpha$ blobs populate a region of our Universe that contains three
filaments. These filaments intersect right where Steidel et
al.\cite{Steidel98} found their enhanced population of
LBGs. Therefore, the survey volume that was probed by \citet{Matsuda04} is
overdense, but not as much as the `sub-volume' that was probed originally by \citet{Steidel00}. Indeed, \cite{Matsuda05} found that the number
density of Ly$\alpha$ emitters in the filaments -- and therefore the
volume originally probed by \citet{Steidel98,Steidel00}-- is approximately two to three times higher than the average number density within their entire volume. If
we assume that the LBG overdensity is suppressed by a similar factor,
then this implies that the overall survey volume of \citet{Matsuda04}
contains $\sim (6.0\pm 1.2)/3\sim 2.0 \pm 0.4$ more LBGs than average.

LBGs populate dark matter halos of mass $m=10^{11.8}M_{\odot}$
\citep{Adelberger05}, which are biased tracers of the overall mass
density field. Therefore, we convert the LBG overdensity, $\delta_{\rm
LBG}=1.0\pm 0.4$ into an overall matter overdensity through
$\delta_{\rm LBG}=b_{\rm LBG}\delta_{\rm M}$, in which $b_{\rm
LBG}\sim 3.0$ is the linear bias parameter of dark matter halos
hosting the LBGs \citep{SMT}. We therefore find that $\delta_{\rm
M}=0.3\pm 0.1$, which corresponds to $\delta_{\rm M}=(3 \pm 1)\times
\sigma(V_{\rm s})$. This result corresponds to the range of
overdensities that was used\footnote{\citet{Lidz} compute the
differential probability that a point is at an overdensity $\delta_R$
when smoothed on a scale $R$, given that it is at an overdensity
$\delta_r$ when smoothed on a smaller scale $r$,
$P(\delta_R,\sigma_R|\delta_r,\sigma_r)$ (see their Eq.~4). For a bias
parameter $b_{\rm LBG}=3.0$ we find that the mass overdensity within
the survey volume of \citet{Steidel98} is $\delta_{\rm M,r}=\frac{\delta_{\rm LBG}}{b_{\rm LBG}}=1.7\pm 0.4=(6.0\pm
1.5)\times \sigma(V_{\rm s})$. Here, the subscripts 'r' and
'R' refer to the smaller and larger survey volumes, respectively. Using the expression
of \citet{Lidz} we find that $\delta_{\rm M,r}=(6.0\pm 1.5)\times
\sigma(V_{\rm s})$ translates to $\delta_{\rm M,R}=(3.3\pm 0.8)\times
\sigma(V_{\rm S})$. This procedure therefore gives us an estimate of
the overdensity that is consistent with our estimate based on the
number of observed LBGs and Ly$\alpha$ emitting galaxies.} in this
paper.


\begin{thebibliography}{14}
\expandafter\ifx\csname natexlab\endcsname\relax\def\natexlab#1{#1}\fi

\bibitem[Adams(1975)]{Adams75} Adams, T.~F.\ 1975, \apj, 201,  350

\bibitem[Adams et al.(2009)]{Adams08} Adams, J.~J., Hill,  G.~J., \&
MacQueen, P.~J.\ 2009, \apj, 694, 314

\bibitem[Adelberger et al.(2005)]{Adelberger05} Adelberger, K.~L.,
Steidel, C.~C., Pettini, M., Shapley, A.~E., Reddy, N.~A.,  \& Erb,
D.~K.\ 2005, \apj, 619, 697

\bibitem[Agertz et al.(2009)]{Agertz09} Agertz, O., Teyssier, 
R., \& Moore, B.\ 2009, \mnras, 397, L64 

\bibitem[Barkana 
\& Loeb(2001)]{2001PhR...349..125B} Barkana, R., \& Loeb, A.\ 2001, \physrep, 349, 125 

\bibitem[Barkana  \& Loeb(2004)]{BL04} Barkana, R., \& Loeb, A.\ 2004,
\apj, 609, 474

\bibitem[Bassett 
\& Woodward(1995)]{BW95} Bassett, G.~M., \& Woodward, P.~R.\ 1995, Journal of Fluid Mechanics, 284, 323 

\bibitem[Basu-Zych \& Scharf(2004)]{Basu04} Basu-Zych, A., \&  Scharf,
C.\ 2004, \apjl, 615, L85

\bibitem[Birnboim \& Dekel(2003)]{Birnboim03} Birnboim, Y., \&  Dekel,
A.\ 2003, \mnras, 345, 349

\bibitem[Birnboim et al.(2007)]{B07} Birnboim, Y., Dekel,  A., \&
Neistein, E.\ 2007, \mnras, 380, 339

\bibitem[Bond et al.(1991)]{Bond91} Bond, J.~R., Cole, S.,
Efstathiou, G., \& Kaiser, N.\ 1991, \apj, 379, 440

\bibitem[Bonilha et al.(1979)]{Bo79} Bonilha, J.~R.~M.,  Ferch, R.,
Salpeter, E.~E., Slater, G.,  \& Noerdlinger, P.~D.\ 1979, \apj, 233,
649


\bibitem[Bunker et  al.(2003)]{Bunker03} Bunker, A., Smith, J.,
Spinrad, H., Stern, D., \& Warren, S.\ 2003, \apss, 284, 357

\bibitem[Chapman et al.(2001)]{Chapman01} Chapman, S.~C., Lewis,
G.~F., Scott, D., Richards, E., Borys, C., Steidel, C.~C., Adelberger,
K.~L., \& Shapley, A.~E.\ 2001, \apjl, 548, L17

\bibitem[Chapman et al.(2004)]{Chapman04} Chapman, S.~C., Scott,  D.,
Windhorst, R.~A., Frayer, D.~T., Borys, C., Lewis, G.~F.,  \& Ivison,
R.~J.\ 2004, \apj, 606, 85

\bibitem[Dekel  \& Birnboim(2006)]{BD06} Dekel, A., \& Birnboim, Y.\
2006, \mnras, 368, 2

\bibitem[Dekel  \& Birnboim(2008)]{BD08} Dekel, A., \& Birnboim, Y.\
2008, \mnras, 383, 119

\bibitem[Dekel et al.(2009)]{Dekel08} Dekel, A., et al.\ 2009,  \nat,
457, 451

\bibitem[Dey et al.(2005)]{Dey05} Dey, A., et al.\ 2005,  \apj, 629,
654

\bibitem[Dijkstra et al.(2006)]{D06} Dijkstra, M., Haiman,  Z., \&
Spaans, M.\ 2006, \apj, 649, 14
\bibitem[Dijkstra et al.(2007)]{IGM} Dijkstra, M., Lidz,  A., \&
Wyithe, J.~S.~B.\ 2007, \mnras, 377, 1175
\bibitem[Dijkstra  \& Loeb(2008)]{DL08} Dijkstra, M., \& Loeb, A.\
2008, \mnras, 386, 492
\bibitem[Dijkstra(2009)]{D09} Dijkstra, M.\ 2009, \apj,  690, 82

\bibitem[Dijkstra \& Westra(2009)]{DW09} Dijkstra, M., Westra, E.,
\mnras in press

\bibitem[Fall  \& Rees(1985)]{FR85} Fall, S.~M., \& Rees, M.~J.\ 1985,
\apj, 298, 18
\bibitem[Fardal et al.(2001)]{Fardal01} Fardal, M.~A., Katz, N.,
Gardner, J.~P., Hernquist, L., Weinberg, D.~H., \& Dav{\'e}, R.\ 2001,
\apj, 562, 605
\bibitem[Faucher-Gigu{\`e}re et al.(2008a)]{Claude}
Faucher-Gigu{\`e}re, C.-A., Lidz, A., Hernquist, L.,  \& Zaldarriaga,
M.\ 2008a, \apjl, 682, L9

\bibitem[Faucher-Gigu{\`e}re et al.(2008b)]{Claude2}
Faucher-Gigu{\`e}re, C.-A., Prochaska, J.~X., Lidz, A., Hernquist, L.,
\& Zaldarriaga, M.\ 2008b, \apj, 681, 831

\bibitem[Furlanetto et al.(2005)]{Fur05} Furlanetto, S.~R.,  Schaye,
J., Springel, V., \& Hernquist, L.\ 2005, \apj, 622, 7

\bibitem[Geach et al.(2005)]{Geach05} Geach, J.~E., et al.\  2005,
\mnras, 363, 1398

\bibitem[Geach et al.(2007)]{Geach07} Geach, J.~E., Smail, I.,
Chapman, S.~C., Alexander, D.~M., Blain, A.~W., Stott, J.~P.,  \&
Ivison, R.~J.\ 2007, \apjl, 655, L9

\bibitem[Geach et al.(2009)]{Geach09} Geach, J.~E., et al.\ 
2009, \apj, 700, 1 

\bibitem[Gao et al.(2008)]{Gao08} Gao, L., Navarro, J.~F.,  Cole, S.,
Frenk, C.~S., White, S.~D.~M., Springel, V., Jenkins, A.,  \& Neto,
A.~F.\ 2008, \mnras, 387, 536

\bibitem[Giodini et al.(2009)]{fbar} Giodini, S., et al.\  2009,
arXiv:0904.0448

\bibitem[Haiman et al.(2000)]{HS00} Haiman, Z., Spaans, M.,  \&
Quataert, E.\ 2000, \apjl, 537, L5

\bibitem[Haiman  \& Rees(2001)]{HR01} Haiman, Z., \& Rees, M.~J.\
2001, \apj, 556, 87
\bibitem[Harrington(1973)]{H73} Harrington, J.~P.\ 1973,  \mnras, 162,
43
\bibitem[Hill et al.(2008)]{Hill08} Hill, G.~J., et al.\ 2008,
Astronomical Society of the Pacific Conference Series, 399, 115
\bibitem[Hopkins et al.(2007)]{H07} Hopkins, P.~F., Lidz,  A.,
Hernquist, L., Coil, A.~L., Myers, A.~D., Cox, T.~J.,  \& Spergel,
D.~N.\ 2007, \apj, 662, 110

\bibitem[Hui  \& Gnedin(1997)]{Hui97} Hui, L., \& Gnedin, N.~Y.\ 1997,
\mnras, 292, 27

\bibitem[Izotov et al.(1997)]{Helium} Izotov, Y.~I., Thuan,  T.~X., \&
Lipovetsky, V.~A.\ 1997, \apjs, 108, 1

\bibitem[Jimenez  \& Haiman(2006)]{JH06} Jimenez, R., \& Haiman, Z.\
2006, \nat, 440, 501

\bibitem[Katz  \& Gunn(1991)]{KG91} Katz, N., \& Gunn, J.~E.\ 1991,
\apj, 377, 365

\bibitem[Katz et al.(1996)]{1996ApJS..105...19K} Katz, N., Weinberg, D.~H., 
\& Hernquist, L.\ 1996, \apjs, 105, 19 

\bibitem[Keel et al.(1999)]{1999AJ....118.2547K} Keel, W.~C., Cohen, S.~H., 
Windhorst, R.~A., \& Waddington, I.\ 1999, \aj, 118, 2547 

\bibitem[Kere{\v s} et al.(2005)]{Keres05} Kere{\v s}, D.,  Katz, N.,
Weinberg, D.~H., \& Dav{\'e}, R.\ 2005, \mnras, 363, 2

\bibitem[Kere{\v s} et al.(2009)]{Keres08} Kere{\v s}, D.,  Katz, N.,
Fardal, M., Dav{\'e}, R.,  \& Weinberg, D.~H.\ 2009, \mnras, 395, 160

\bibitem[Komatsu et al.(2009)]{Komatsu08} Komatsu, E., et al.\  2009,
\apjs, 180, 330

\bibitem[Lidz et al.(2007)]{Lidz} Lidz, A., McQuinn, M.,  Zaldarriaga,
M., Hernquist, L., \& Dutta, S.\ 2007, \apj, 670, 39

\bibitem[Madau(1995)]{Madau95} Madau, P.\ 1995, \apj, 441, 18

\bibitem[Makino et al.(1998)]{Makino} Makino, N., Sasaki, S.,  \&
Suto, Y.\ 1998, \apj, 497, 555
\bibitem[Martini(2004)]{M04} Martini, P.\ 2004, Coevolution  of Black
Holes and Galaxies, 169
\bibitem[Matsuda et al.(2004)]{Matsuda04} Matsuda, Y., et al.\  2004,
\aj, 128, 569
\bibitem[Matsuda et al.(2005)]{Matsuda05} Matsuda, Y., et al.\  2005,
\apjl, 634, L125

\bibitem[Matsuda et al.(2006)]{Matsuda06} Matsuda, Y., Yamada,  T.,
Hayashino, T., Yamauchi, R., \& Nakamura, Y.\ 2006, \apjl, 640, L123

\bibitem[Matsuda et al.(2007)]{Matsuda07} Matsuda, Y., Iono, D.,
Ohta, K., Yamada, T., Kawabe, R., Hayashino, T., Peck, A.~B., \&
Petitpas,  G.~R.\ 2007, \apj, 667, 667

\bibitem[Miles(1957)]{Miles57} Miles, J.~W.\ 1957, Acoustical 
Society of America Journal, 29, 226 

\bibitem[Mori et al.(2004)]{Mori04} Mori, M., Umemura, M., \& Ferrara,
A.\ 2004, \apjl, 613, L97
\bibitem[Neufeld(1990)]{N90} Neufeld, D.~A.\ 1990, \apj,  350, 216
\bibitem[Nilsson et al.(2006)]{Nilsson06} Nilsson, K.~K., Fynbo,
J.~P.~U., M{\o}ller, P., Sommer-Larsen, J., \& Ledoux, C.\ 2006, \aap,
452,  L23

\bibitem[Navarro et al.(1997)]{NFW} Navarro, J.~F., Frenk,  C.~S., \&
White, S.~D.~M.\ 1997, \apj, 490, 493

\bibitem[Ocvirk et al.(2008)]{Ocvirk08} Ocvirk, P., Pichon, C.,  \&
Teyssier, R.\ 2008, \mnras, 390, 1326
\bibitem[Osterbrock(1989)]{Os89} Osterbrock, D.~E.\ 1989,  Research
supported by the University of California, John Simon Guggenheim
Memorial Foundation, University of Minnesota, et al.~Mill Valley, CA,
University Science Books, 1989, 422 p.,

\bibitem[Press  \& Schechter(1974)]{PS74} Press, W.~H., \& Schechter,
P.\ 1974, \apj, 187, 425

\bibitem[Rees  \& Ostriker(1977)]{RO77} Rees, M.~J., \& Ostriker,
J.~P.\ 1977, \mnras, 179, 541

\bibitem[Rees(1989)]{Rees89} Rees, M.~J.\ 1989, \mnras, 239,  1P
\bibitem[Rybicki  \& Lightman(1979)]{RL79} Rybicki, G.~B., \&
Lightman, A.~P.\ 1979, New York, Wiley-Interscience, 1979.~393 p.,
\bibitem[Scharf et al.(2003)]{Scharf03} Scharf, C., Smail, I.,
Ivison, R., Bower, R., van Breugel, W.,  \& Reuland, M.\ 2003, \apj,
596, 105
\bibitem[Shapley et al.(2003)]{S03} Shapley, A.~E.,  Steidel, C.~C.,
Pettini, M., \& Adelberger, K.~L.\ 2003, \apj, 588, 65
\bibitem[Smith \& Jarvis(2007)]{Smith07} Smith, D.~J.~B., \&  Jarvis,
M.~J.\ 2007, \mnras, 378, L49

\bibitem[Saito et al.(2006)]{Saito06} Saito, T., Shimasaku, K.,
Okamura, S., Ouchi, M., Akiyama, M., \& Yoshida, M.\ 2006, \apj, 648,
54

\bibitem[Saito et al.(2008)]{Saito08} Saito, T., Shimasaku, K.,
Okamura, S., Ouchi, M., Akiyama, M., Yoshida, M.,  \& Ueda, Y.\ 2008,
\apj, 675, 1076

\bibitem[Schaye et al.(2003)]{Schaye03} Schaye, J., Aguirre, A.,  Kim,
T.-S., Theuns, T., Rauch, M.,  \& Sargent, W.~L.~W.\ 2003, \apj, 596,
768

\bibitem[Shapley et al.(2003)]{Shapley03} Shapley, A.~E.,  Steidel,
C.~C., Pettini, M., \& Adelberger, K.~L.\ 2003, \apj, 588, 65

\bibitem[Sheth  \& Tormen(1999)]{ST} Sheth, R.~K., \& Tormen, G.\
1999, \mnras, 308, 119

\bibitem[Sheth et al.(2001)]{SMT} Sheth, R.~K., Mo, H.~J.,  \& Tormen,
G.\ 2001, \mnras, 323, 1

\bibitem[Shull  \& van Steenberg(1985)]{SV85} Shull, J.~M., \& van
Steenberg, M.~E.\ 1985, \apj, 298, 268
\bibitem[Smith \& Jarvis(2007)]{Smith07} Smith, D.~J.~B., \&  Jarvis,
M.~J.\ 2007, \mnras, 378, L49
\bibitem[Smith et al.(2008a)]{Smith08a} Smith, D.~J.~B., Jarvis,
M.~J., Lacy, M., \& Mart{\'{\i}}nez-Sansigre, A.\ 2008a, \mnras, 389,
799

\bibitem[Smith et al.(2009)]{Smith08b} Smith, D.~J.~B., Jarvis,
M.~J., Simpson, C., \& Mart{\'{\i}}nez-Sansigre, A.\ 2009, \mnras,
393, 309

\bibitem[Somerville et al.(2004)]{S04} Somerville, R.~S.,  Lee, K.,
Ferguson, H.~C., Gardner, J.~P., Moustakas, L.~A.,  \& Giavalisco, M.\
2004, \apjl, 600, L171

\bibitem[Steidel et al.(1998)]{Steidel98} Steidel, C.~C.,  Adelberger,
K.~L., Dickinson, M., Giavalisco, M., Pettini, M.,  \& Kellogg, M.\
1998, \apj, 492, 428

\bibitem[Steidel et al.(2000)]{Steidel00} Steidel, C.~C.,  Adelberger,
K.~L., Shapley, A.~E., Pettini, M., Dickinson, M.,  \& Giavalisco, M.\
2000, \apj, 532, 170

\bibitem[Steidel et al.(2003)]{Steidel03} Steidel, C.~C.,  Adelberger,
K.~L., Shapley, A.~E., Pettini, M., Dickinson, M.,  \& Giavalisco, M.\
2003, \apj, 592, 728

\bibitem[Taniguchi \& Shioya(2000)]{Taniguchi00} Taniguchi, Y., \&
Shioya, Y.\ 2000, \apjl, 532, L13

\bibitem[Villar-Mart{\'{\i}}n(2007)]{V07}  Villar-Mart{\'{\i}}n, M.\
2007, New Astronomy Review, 51, 194

\bibitem[Weidinger et al.(2004)]{Weidinger04} Weidinger, M.,
M{\o}ller, P., \& Fynbo, J.~P.~U.\ 2004, \nat, 430, 999

\bibitem[Yang et al.(2006)]{Yang06} Yang, Y., Zabludoff,  A.~I.,
Dav{\'e}, R., Eisenstein, D.~J., Pinto, P.~A., Katz, N., Weinberg,
D.~H., \& Barton, E.~J.\ 2006, \apj, 640, 539

\bibitem[Yang et al.(2009)]{Yang08} Yang, Y., Zabludoff, A.,
Tremonti, C., Eisenstein, D., \& Dav{\'e}, R.\ 2009, \apj, 693, 1579

\end{thebibliography}
\end{document}